\documentclass[final,5p,twocolum]{elsarticle}

\usepackage{graphics}
\usepackage{graphicx}
\usepackage{epsfig}
\usepackage{amssymb}
\usepackage{amsthm}
\usepackage{lineno}
\usepackage{setspace}
\usepackage{lscape}
\usepackage{xcolor}
\usepackage{amsmath}
\usepackage{makecell}
\usepackage{hyperref}

\journal{Applied Thermal Engineering}

\begin{document}

\begin{frontmatter}

\title{Combined active-passive heat transfer control using slotted fins and oscillation in turbulent flow: the cases of single cylinder and tube banks}
\author[1]{Haleh Soheibi\corref{cor2}}
\author[2]{Zahra Shomali\corref{cor2}}
\cortext[cor2]{These authors contributed equally to this work}
\author[1]{Jafar Ghazanfarian\corref{cor1}}
\cortext[cor1]{Corresponding author, Tel.: +98(24) 3305 4142.}
\ead{j.ghazanfarian@znu.ac.ir}
\address[1]{Mechanical Engineering Department, Faculty of Engineering, University of Zanjan, P.O. Box 45195-313, Zanjan, Iran.}
\address[2]{Department of Physics, Faculty of Basic Sciences, Tarbiat Modares University, Tehran, Iran}

\begin{abstract}
In heat transfer augmenting methods such as radial fins, the heat transfer enhancement commonly leads to the drag force increment. In the present paper, slots are inserted over the fins to simultaneously reduce the drag coefficient. Turbulent convection heat transfer around a cylinder, as well as oscillating bundle of tubes including the slotted radial fins have been investigated. The governing equations are solved in two-dimension utilizing OpenFOAM software based on k-$\omega$ SST closure model. The cases with various slot location, slot width, the number of the slots, the fin height, and oscillation frequencies are examined. In all cases, the Reynolds number is taken to be equal to 5000. Presence of three slots on the fins reduces the drag coefficient by 23$\%$ and augments the Nusselt number by 76$\%$. In order to enhance the heat transfer from the bundle of tube, oscillation of tubes and utilization of the slotted fins are applied. The optimum situation occurs for a sample with the oscillating third column that shows 3$\%$ increment in heat transfer relative to that of the fixed case. This is while adding the slotted fins to the oscillating tube bank increases the heat transfer up to 2.5 times.
\end{abstract}


\end{frontmatter}

\section{Introduction}
Forced convection around cylinder is a momentous engineering issue that is widely employed in industries, including heat exchangers, heaters, refrigerators, nuclear reactors, electronic devices, and even suspended bridge cables. Over the past decades, researchers have been looking for the methods to improve the heat transfer rate to design cost-effective systems, and also reduce the energy consumption. In general, the methods for enhancement of heat transfer are placed into three groups: active methods, passive methods, and combined methods. Active strategies are the methods in which the heat transfer improvement occurs by connecting the system to an external energy source. Passive methods require no external energy source, and usually enhance the heat transfer coefficient by modifying the geometry to effectively disturb the flow or additives to change the physical characteristics of the fluid. Surface oscillation, fluid oscillation, suction and fluid injection are distinguished active methods. The most frequently used passive methods include utilizing the expanded surfaces or fins, surface roughening in micro/nanoscale, and the addition of nano-particles to the fluid. In combined techniques, both active and passive methods are used simultaneously to enhance the heat transfer.

Increasing the heat transfer coefficient of heat exchangers by the three mentioned methods, also augments the frictional losses. As a result, there exists a need to higher pumping power to accelerate the fluid. To optimize the performance of heat exchangers, both factors should be contemplated. Among the methods for augmenting the heat transfer coefficient, surface oscillation is one of the most common ones. The heat exchangers that use this method are called oscillating heat exchangers~\cite{Ghazanfarian2017}. During oscillation, the lock-on phenomenon occurs, which is due to the synchronization of the excitation frequency with the vortex shedding natural frequency. Attachment of different types of fins is also a common passive method that extends the surface of heat exchanging. It also affects the heat transfer and the friction by modifying the flow pattern. Many theoretical, numerical, and experimental researches have been devoted to investigation of effect of parameters such as the amplitude and frequency of oscillations, the number of fins and their length on the heat transfer rate and force drag coefficient.

Griffin \emph{et al.}~\cite{GRIFFIN1971} experimentally studied the flow around oscillating cylinder at low Reynolds number. It has been demonstrated that when the cylinder oscillates with the frequency equal or near to the vortex shedding frequency, the length of the vortex formation region is shortened in comparison to that of the fixed cylinder. Also, the flow structure in the downstream vortical region is found to be completely influenced by the amplitude and frequency of excitations. Williamson and Roshko \cite{Williamson1988} investigated a cylinder with cross flow oscillations. They have presented different sketches and also the corresponding modes of vortex shedding. It has been found that when the frequency of the vortex shedding becomes equal to the excitation frequency, the lock-in with specified modes takes place. The different modes of vortex shedding, the lock-on and the non-lock-on regions are determined in their reported wavelength-amplitude plots.

Cheng \emph{et al}.~\cite{Cheng1997} experimentally studied the forced heat transfer around the oscillating cylinder in the range of Re$<$4000. The heat transfer is seen to be notably augmented by making the cylinder oscillate. They attributed such enhancement to the lock-on and also the turbulence effects. They also numerically explored the effect of the lock-on effect on convective heat transport through the oscillating cylinder at Re$<$300~\cite{Cheng1997p2}. They obtained that the heat transfer performance substantially increases in the lock-on region. In 1999, Gau \emph{et al.}~\cite{Gau1999} experimentally investigated thermal transport over a cylinder oscillating in cross-flow at harmonic, subharmonic, superharmonic, and nonharmonic frequencies. The dimensionless frequency defined as the cylinder excitation frequency, $F_e$, to the normal frequency of vortex shedding, $F_n$, equals 0.5, 1, 2, 2.5, 3. It is shown that the lock-in takes place for both $\frac{F_e}{F_N}$=1, and 3, which itself results in augmentation of the heat transfer mechanism. The effect of a cylinder placed in a uniform stream and oscillating about its axis on the forced heat convection at Re$<$40 have been numerically studied in~\cite{Mahfouz1999}.

The lock-on phenomenon also appears in the case of rotationally oscillating cylinder. It is reported that the lock-in occurs at excitation frequencies that are close to the natural frequency. In the lock-on regime, by increasing the frequency, the heat transfer coefficient enhances; while outside this region, the oscillation does not much affect the heat transfer. Further, the transversely oscillating cylinder in a cross flow has been numerically investigated by Fu and Tong \cite{Fu2000}. It has been announced that in the lock-in regime, the flow and thermal fields approach a periodic state. Also, the remarkable increase of the heat transfer in the lock-in regime has been mentioned. In a continuation of the previous works, the influence of oscillation on heat transfer is inquired experimentally at low Reynolds numbers~\cite{Pottebaum2006}. Their results confirmed that at frequencies close to the natural frequency, a significant augmentation in heat transfer happens. The heat transfer coefficient is found to be dependent on several factors such as shortening the length of the vortex formation region, the details of the vortex roll-up procedure in the cylinder wake, and the transverse cylinder velocity.

Kumar \emph{et al.}\cite{Kumar2016} introduced new criteria for the lock-in phenomenon by investigating the flow around an oscillating cylinder at Re=100. They explained that the lock-in occurs when two following conditions are satisfied. The first clause is the existence of the most dominant frequency in the power spectrum of the lift coefficient that matches the frequency of the cylinder oscillation. The second condition states that if other peaks exist in the power spectrum, it should be definitely at super-harmonics of the cylinder oscillation frequency with the value of integral multiples of it. The regime where only the first of the two conditions is satisfied is referred to as the transition regime. All other situations correspond to a non-locked state. Consequently, three flow regimes: the lock-in, transition, and the non-locked can be recognized.

More, the fluid flow passing through an oscillating tube bank in aligned arrangements has been numerically explored~\cite{Ghazanfarian2017}. To augment the heat transfer, the first column of the tubes in the lock-in region is vibrated with different frequencies and amplitudes. Consequently, approximate 25\% augmentation of the overall Nusselt number has been obtained. Furthermore, the effect of oscillating first column on heat transfer enhancement in low-amplitude motion is found to be remarkable. It is recommended to oscillate the first column with low amplitude and the frequency inside the lock-in regime, to reach the maximum conceivable heat transfer with the lowest construction costs.

By adding nanoparticles into the base fluid as well as oscillating the cylinder, the heat transfer rate can be enhanced~\cite{Mousavi2018}. The study considers low Reynolds numbers and also takes the thermodynamical properties, temperature-dependent. It has been acquired that addition of the nanoparticles into the base fluid, increases the nanofluid viscosity and consequently it lessens the vortex shedding frequency. Moreover, the maximum 54.82\% augmentation of the drag coefficient and also 9.66\% enhancement of the heat transfer coefficient due to the transverse oscillation of the cylinder is reported. Furthermore, utilizing both nanofluid and transverse oscillation found to improve the heat transfer up to 30.3 and 26.8\%, respectively, for static and dynamic models.

Besides, the laminar flow around the rotating cylinder with external longitudinal fins, the situation which one confronts in electrical motors and generators, is numerically investigated in~\cite{Murthy1983}. Rotation of the cylinder is found to cause the recirculating flow in the inter-fin space. Further, it has been shown that the local heat transfer coefficients are large near the front face of the fin and also close to the fin tips. Also, the cross-flow forced convection heat transfer has been studied numerically~\cite{Abu-Hijleh2003,Abu-Hijleh2003p2}.

When Re$<$200, the maximum Nusselt number for a given value of fin height occurs for an optimum number of fins such that utilizing more of fins make the Nusselt number decrease. Also, the authors have investigated the effect of using permeable and solid fins. For a similar cylinder configuration, the permeable fins found to provide much higher Nusselt number than that of the traditional solid fins. By increasing the Reynolds number and the fin height, the Nusselt number increases while it decreases with increment of the number of fins. More, using one or two permeable fins results in much higher Nusselt number in comparison to the one obtained for the configuration with eighteen solid fins. Also, utilizing porous fins notably reduces the weight and cost.

Later, the turbulent natural convection between two horizontal concentric cylinders with radial fins has been numerically investigated utilizing the standard k-$\epsilon$ model~\cite{Rahnama2004}. For the number of fins between 2 to 12 and also the Rayleigh number of 10$^6$$<$Ra$<$10$^9$, the streamlines and the temperature contours have been presented. It has been obtained that by increasing the Rayleigh number, the heat transfer rate enhances. Also, the higher fin heights are shown to have a blocking effect on the flow and reduces the Nusselt number. Also the laminar free convection of air around a horizontal cylinder with external longitudinal fins has been numerically investigated~\cite{Haldar2004}.

It has been reported that increasing the Grashof number, the number of fins, and the fin length enhances the heat transfer rate. For a identical fin surface, it is found that larger number of fins with smaller length augments the heat transfer rate. Haldar \emph{et al.} also performed the conjugate numerical computation~\cite{Haldar2007}. They obtained that the fin thickness has the dominant influence on the heat transfer. It is established that adding more fins to the tube gives rise to the enhancement of the heat transfer, if only the fins are thin. The particle image velocimetry (PIV) has been used to study the wake region behind a foamed and a finned cylinder in cross-flow \cite{Khashehchi2014} with focus on turbulence. The measurements have been done for a wide interval of the Reynolds number, 1000$<$Re$<$10000. Also, the influence of the fin pitch ratio on wake development and structural loading specifications of the cylinder with equispaced circular fins have been experimentally accomplished by the PIV technique~\cite{McClure2016}. It is obtained that increasing the number of fins results in augmentation of the mean drag due to raising the skin friction. At a critical value of the fin pitch, coalescence of the boundary layers between adjacent fins causes a notable augmentation of the vortex formation region.

Utilizing the OpenFOAM toolbox, the influence of adding straight fins on thermal and hydraulic properties of the transient heat and fluid flow over a circular cylinder has been studied~\cite{Bouzari2016}. It is obtained that as the fins are lengthened, increment of the mean drag coefficient takes place, while the average Nusselt number decreases. Also, it is concluded that the fin effectiveness is optimum for all Reynolds numbers when the case has four fins. On the other hand, the effect of adding straight circular fins to the cylinder has been experimentally investigated in~\cite{Islam2020}. In this study, the finned cylinders with different diameter ratios, $D_f/D_r$, and the same fin pitch and thickness are investigated. The PIV measurements are performed at the Reynolds number of Re=2.0$\times$10$^4$ correspondent to the sub-critical flow regime. The notable decrease in the recirculation region followed by an amplified primary vortex shedding procedure in the wake of the cylinders are seen. This amplification results in increase of the momentum transport owing to the periodic Reynolds stresses in the wake of the finned cylinders with a higher diameter ratio. Moreover, the pressure at the lee side increases due to the growth of $D_f/D_r$, which itself causes a reduction in the drag force.

There are many studies devoted to the reduction of the drag forces and also suppressing the vortex shedding from the circular cylinders. In 2002, Tsutsui and Igarashi \cite{Tsutsui2002} set a rod upstream of the circular cylinder, to control the flow around a circular cylinder in air-stream for 1.5$\times$10$^4$$<$Re$<$6.2$\times$10$^4$. Existence of two flow patterns: with and without the vortex shedding from the rod has been reported. The flow pattern varies for different values of the rod diameter and position and also the Reynolds number. It has been obtained that the total drag comprising the drag of the rod is 63\% reduced in comparison to that of the single cylinder. In the following, the experimental study of placing a rod upstream of and parallel to the cylinder to control the flow around the cylinder were performed for Re=8.2$\times$10${^4}$ \cite{Wang2006}. The resultant force of the cylinder reduces for spacing between the rod and the cylinder $\alpha$ (the staggered angle of the rod and cylinder) less than 5$^{\circ}$. When $\alpha$=0$ ^{\circ}$ and the ratio of the diameter of the rod relative to the cylinder diameter is equal to 0.5, the maximum drag reduces to 2.34$\%$ of that of the single cylinder.

The effect of utilizing the dual detached splitter plates, with the same length as the cylinder diameter, for a circular cylinder on the drag reduction is numerically studied in \cite{Hwang2007}. One splitter plate is placed upstream of the cylinder, while the other splitter is located in the near-wake region. The stagnation pressure is reduced by the upstream splitter and the downstream one increases the base pressure by tranquilizing the vortex shedding. The combined affect results in a notable drag reduction on the cylinder. Also, the aerodynamic differences between a bare cylinder and a semi-cylindrical roof have been investigated via wind tunnel experiments by using a circular cylinder with splitter \cite{Qiu2014}. It is obtained that a frontal splitter plate creates a postcritical flow at relatively low Reynolds numbers by the generated disturbances.

Later in 2017, Klausmann and Ruck \cite{Klausmann2017} experimentally investigated the flow around cylinder when 3$\times$10$^4<$Re$<$1.4$\times$10$^5$. It has been shown that the thin porous layer on the leeward side of the cylinder, causes the enhancement of the base pressure on the leeward side of the cylinder that leads to a decrease in the drag force and dampens the oscillation amplitudes. The maximum drag reduction of 13.2\% has been measured. Furthermore, the drag suppression for a high-amplitude oscillating cylinder is investigated utilizing different configurations of the detached-splitter plates \cite{Amiraslanpour2017}. It has been obtained that the case with the double upstream splitter has the highest drag reduction. The maximum reduction of the drag force takes place for the lock-on condition at dimensionless oscillation frequency of $f^*$=1.1. The gap size and the plate length are found to be the least and the most effective quantities.

As it is clear, there have been many studies dealing with the oscillating movement of the cylinder and its effects on flow and heat transfer \cite{Ghazanfarian2015,Ghazanfarian2016,Ghazanfarian2019}. Also, the effects of the different parameters such as the various types of fins, including annular and radial ones are investigated. Several methods for the reduction of the drag coefficient of the cylinder, suchlike utilizing different kinds of splitters or pipes have been investigated.

In the present study, the effect of existence of slot on a radial fin on reducing the drag coefficient is investigated. As mentioned in Ghazanfarian and Bouzari \cite{Bouzari2016}, where the fluid flow over a circular cylinder with radial fin has been investigated, a high drag coefficient appears for a cylinder with four radial fins. However, to the best of our knowledge, there have been no study aiming at reduction of the drag coefficient of the cylinder with radial fin in cross-flow. The present scrutiny will focus on such effect. Here, we also explore a cylinder with four fins and slots considered in two vertical fins. The fluid can flow through the slots and affect the drag coefficient and the heat transfer by changing the flow structure. The effects of different parameters such as the position of the slot on the radial fin, the width of the slot, the number of slots, and the height of the slotted fin will be also investigated. Further, the effect of oscillation of different columns of the bundle, and also putting up the radial slotted fin on all tubes on enhancement of the heat transfer in the staggered tube bank will be studied.

This paper is organized as follows. In Sec. \ref{geo}, the details of geometry, the governing equations, and assumptions will be presented. The details of the numerical method are discussed in Sec. \ref{method}. Validation of the results for four different cases of oscillating and fixed cylinders is presented in Sec. \ref{validation}. Then, a precise discussion about the obtained results with adequate engineering criteria is presented in Sec. \ref{results}. First, the results for simulating flow around the cylinder with slotted fin and the effects of geometric parameters of the slots on flow and the heat transfer are reported. Second, the slotted fin placed on the oscillating tube bank and impacts of the oscillation and the fin on heat transfer from the staggered bundle of tubes are examined. Finally, Sec. \ref{con} presents concluding remarks.

\section{Geometry and problem description}
\label{geo}
\begin{figure}
 	\centering
 	\includegraphics[width=\columnwidth]{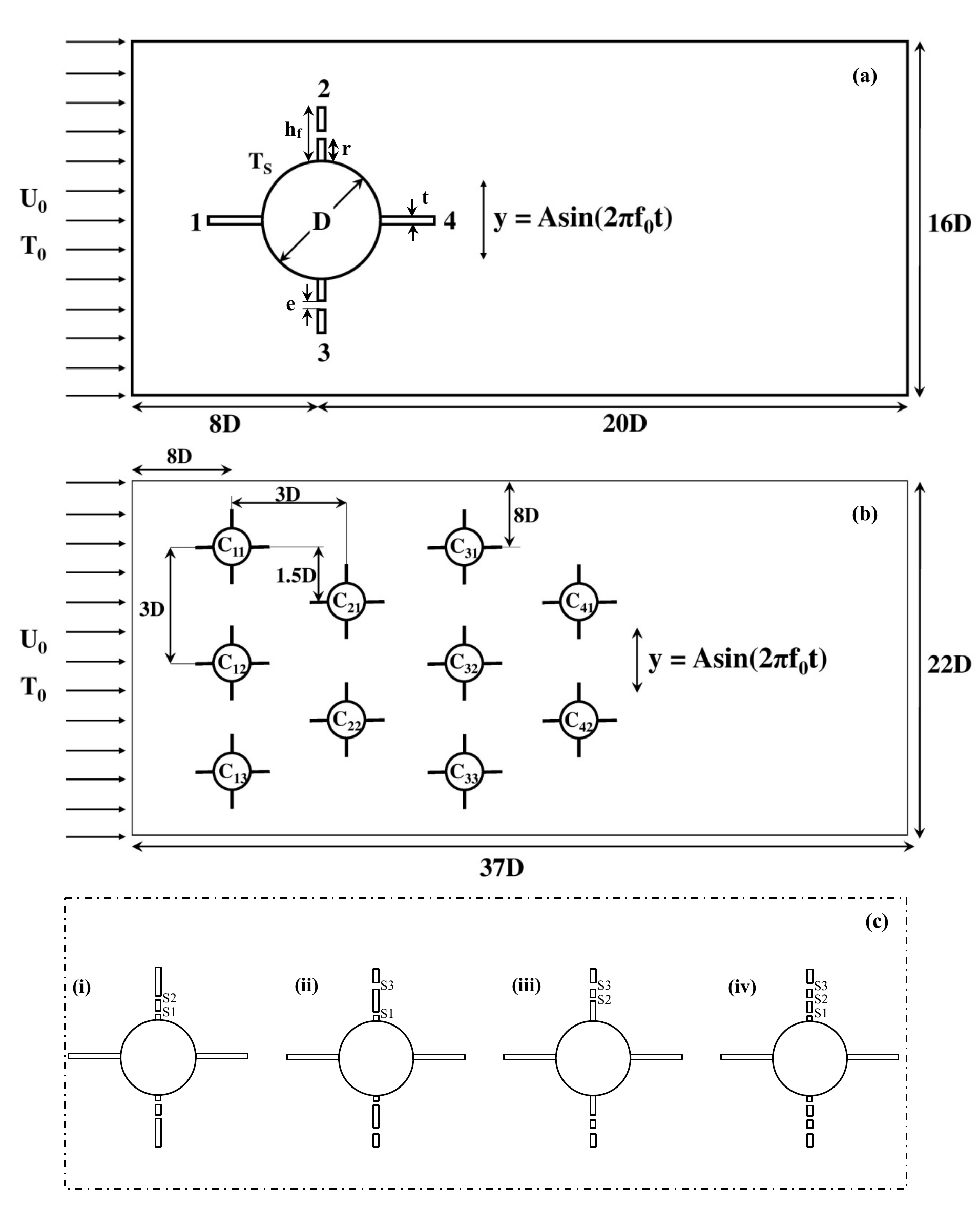}
 	\caption{The studied geometries: (a) a finned tube inside the channel, (b) bundle of finned tubes inside the channel, (c) two/three-slotted cases.}
 	\label{Fig11}
 \end{figure}

Schematic geometry of the cylinder with slotted fin has been shown in Fig. \ref{Fig11}. A cylinder with diameter D and four fins, and also with slots on two vertical fins has been placed in cross-flow. The flow with velocity U$_0$ enters the channel from the left at temperature T$_0$ and passes the finned cylinder with surface temperature of T$_s$.

The distance of the center of cylinder from the channel entrance and the walls around the channel is taken to be 8 times the cylinder diameter. Also, the distance of the cylinder from the channel exit equals 20 times of D. Here, the connected fins are called as number 1 to number 4. The oscillatory motion of the cylinder in normal-to-flow direction obeys the following relation:
\begin{equation}
	\label{yA}
	y=A\ sin \ (2\pi f_0 t),
\end{equation}
where A is the amplitude of oscillation and f$_0$ is the excitation frequency. A$^*$ is the dimensionless amplitude scaled with respect to the diameter. In the present case, A$^*$ is taken to be 0.2, 0.4, 0.6. The dimensionless excitation frequency is the ratio of excitation frequency to the vortex shedding frequency. In this study, all the oscillations occur in frequencies near to the vortex shedding frequency and consequently f$^*$ is close to 1. The dimensionless parameters are:
\begin{equation}
	A^{*}=\frac{A}{D}; \ f*=\frac{f_0}{f_s}; \ Re=\frac{U_0 D}{\nu}; \ St=\frac{f_s D}{U_0}; \ Nu=\frac{hD}{k}= \nonumber
\end{equation}
\begin{equation}
	\frac{\partial T/\partial n \ D}{T_s-T_0};  \ Pr=\frac{\nu}{\alpha};  \  \xi=\frac{Q_{Finned}}{Q_{No-Fin}}; \ H=\frac{h_f}{R}; \ r^*=\frac{r}{D}; \nonumber
\end{equation}
\begin{equation}
	e^*=\frac{e}{D};  \ t^*=\frac{t}{D};  \  T^*=\frac{T-T_0}{T_s-T_0}.	
\end{equation}
The Reynolds number, $Re=\frac{U_0 D}{v}$, equals 5,000. Here, $v$ is the kinematic viscosity of the fluid. The dimensionless vortex shedding frequency is called the Strouhal number, St. In the definition of the Nusselt number, Nu, $h$ and $k$ are respectively, the heat transfer coefficient and the thermal conductivity of the fluid. Also, $\partial T/\partial n$ is the vertical-to-surface temperature gradient. Herein, the Prandtl number, which is the ratio of $v$ to thermal diffusivity $\alpha$ is considered to be 0.71.

The dimensionless number $\psi$, is introduced as the Fin-Effectiveness to investigate the performance of the fin. This number is defined as the ratio of the Heat flux rate from the tube with fin to the heat flux from the tube without fin (naked tube). The inset of Fig.~\ref{Fig11}(a), demonstrates the geometrical parameters of the fin and the slots on it. The height of the fin, h$_f$, is scaled to the radius of the cylinder. H is the dimensionless length of the fin scaled to the $D$. In the current study, H is contemplated 0.65, 0.75, 0.85. $r$ is the distance of the center of slot from the connection point of the fin to the tube. The location of the slot is described by the dimensionless number $r^*$. The problem is investigated for the slot being in three different positions of 0.05, 0.15, 0.25. The other important non-dimensional parameter is the ratio of the width of the slot to the diameter, $e^*$, that is taken to be 0.01, 0.02, 0.06, 0.4. Further, the scaled thickness of the Fin, $t^*$, is set to 0.03.

The thermal resistance of the fin is ignored due to its narrowness and high thermal conductivity. The temperature at the surface of the tube and the fins is a constant equal to T$_s$. The non-dimension temperature, T$_*$, varies between zero and one. T$^*=0$ is the non-dimensional temperature of the free-stream and T$^*=1$ is that of the cylinder surface. The next studied geometry is the staggered bundle of tubes with slotted fins, see Figs. \ref{Fig11}(b) and (c). As it is seen, the tubes in the tube bundle are placed in four columns. In \cite{Ghazanfarian2017}, where an inline tube bank has been investigated, the maximum Nu/C$_d$ is found to belongs the case for which the distance between the center of the tubes is three times the cylinders’ diameter. Herein, as it is seen in Fig.~\ref{Fig11}(c), the vertical and horizontal distance between two adjacent cylinders is taken to be three times the diameter of the tube. Further, the distance of the inlet boundary from the center of the tubes of first column, the upper wall from the center first column’s tubes, and also the lower wall from the tubes of the third column, are the same and equal to the eight times of the cylinder diameter. The last column is 20D far from the outflow boundary. The cylinders are named as C$_{i,j}$ with $i$ and $j$, respectively, denoting the number of the column and the row. For instance, C$_{32}$ refers to the tube located in the third column and the second row. Each column can oscillate perpendicular to the flow according to Eq.~(\ref{yA}).
\section{Governing equations and closure}
\label{method}
The governing equations are the time-averaged version of the mass, momentum, and energy conservation equations,
\begin{equation}
	\label{2-14}
 \begin{aligned}		
\frac{\partial \bar{u_i}}{\partial x_i}=0
 \end{aligned}
\end{equation}
\begin{equation}
		\label{2-15}
\frac{\partial \bar{u_i}}{\partial t}+\frac{\partial \bar{u_i}}{\partial x_j}=-\frac{1}{\rho}\frac{\partial\bar{p}}{\partial x_i}+\nu \frac{\partial^2\bar{u_i}}{\partial x^{2}_{j}}-\frac{\partial \bar{u’_i}u’_j}{\partial x_j}.	
	\end{equation}
\begin{equation}
		\label{2-16}
\left.\begin{aligned}
&\frac{\partial{\bar{T}}}{\partial t}+\frac{\partial{\bar{T}}}{\partial x_j}=\alpha \frac{\partial^2 \bar{T}}{\partial x^{2}_{j}}.&	
\end{aligned}\right.
	\end{equation}
Here, $\rho$, $v$, and $\alpha$ are, respectively, the fluid density, the fluid kinematic viscosity, and the diffusion coefficient. Moreover, $\bar{u_l}$ is the time-averaged velocity while $u’_{l}$ denotes the fluctuations of velocity.

The boundary conditions are the no-slip and no-temperature jump, and zero gradient pressure at the inlet boundary. The K-$\omega$ SST model as a two-equation model, has been used to simulate the flow. This model, proposed by Menter in 1993, is a combination of the K-$\omega$ and $K-\epsilon$ models. The K-$\omega$ model, which is more efficient than the K-$\epsilon$ model in solving the near wall flow, is more appropriate for the study of boundary-layer suffering separation undergoing intense adverse pressure gradients. This turbulent model strongly depends on the boundary conditions of the free-stream at inlet. On the other side, the K-$\epsilon$ turbulent model, is suitable for resolving the shear flow like vortex shedding downstream of the bluff bodies and is less sensitive to the inlet BCs. This model is not adequate for near-wall flow where the local Reynolds number is low. This model presents more accurate predictions relative to the other RANS models and, consequently, it has attracted more attentions in recent years. Here, the RANS discretized equations of turbulent flow are solved using the finite volume method (FVM) and the opensource OpenFOAM software. The following relations are the governing equations in incompressible Newtonian form,
\begin{eqnarray}
			\label{2-17}
\nonumber \frac{D}{Dt}(\rho\omega)&=&\nabla.(\rho D_\omega \nabla \omega)+\frac{\rho \lambda G}{\nu}- \frac{2}{3}\rho\gamma \\
\nonumber \omega(\nabla.u)&-&\rho\beta \omega^2  -\rho(F_{1}-1)CD_{k\omega}+S_{\omega} \\
\nonumber \frac{D}{Dt}(\rho k)&=&\nabla.(\rho D_k \nabla k)+\rho G-\frac{2}{3}\rho k (\nabla.u)-\rho \beta^*\omega k+S_{k}\\
	\end{eqnarray}
Here, $k$ is the turbulent kinetic energy and $\omega$ is the rate of dissipation, which when solved, yields the Reynolds stress tensor. The used solver is “themalPimpleDyMFoam” that is created by adding the energy equation to the “pimpleDyMFoam” solver. To discretize the time derivative terms, the Euler method is used. Also, the advection terms and the diffusion and gradient terms are, respectively, calculated via the Gauss-linear upwind and the Gauss-linear schemes.

\begin{figure}
	\centering
	\includegraphics[width=0.9\columnwidth]{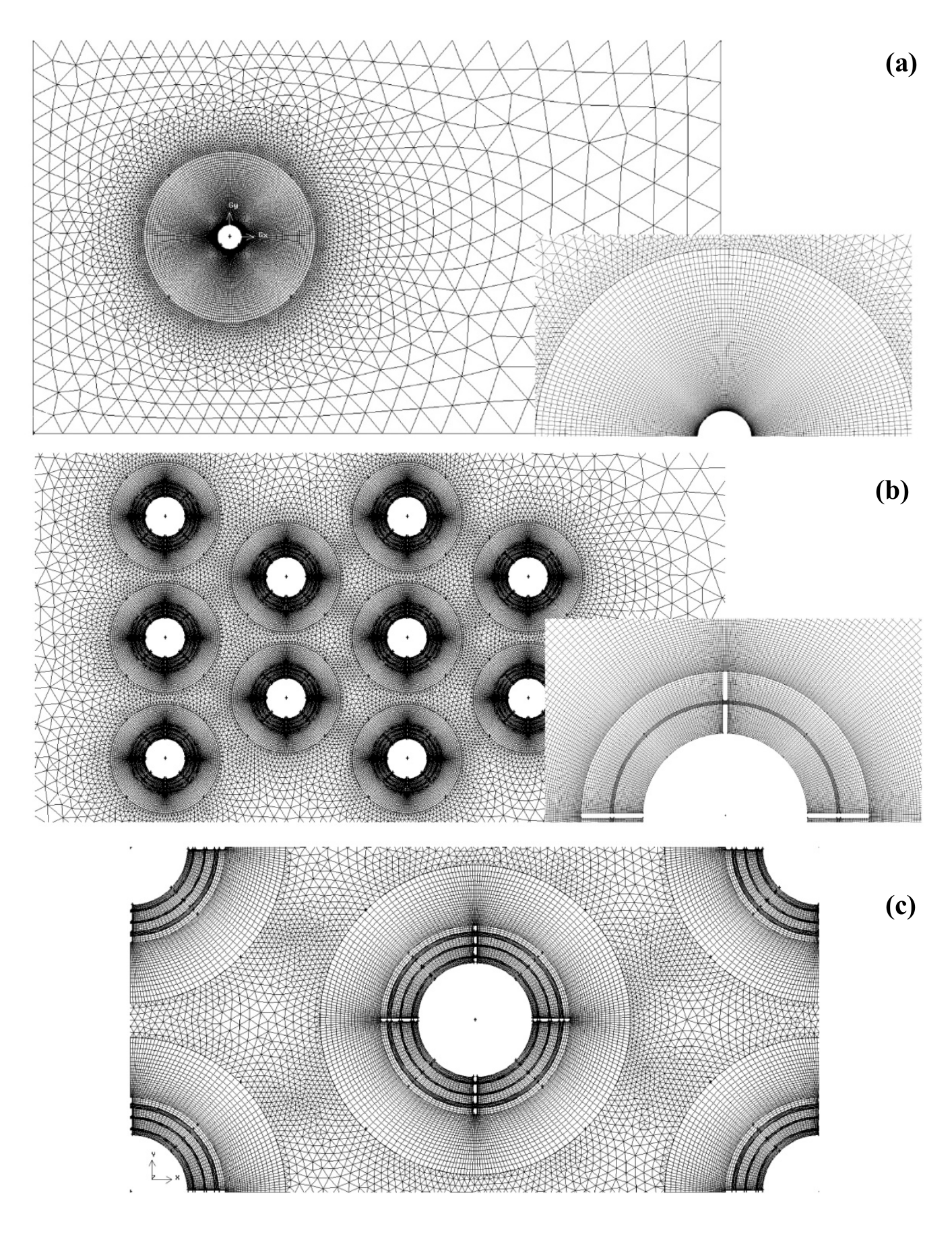}
	\caption{Unstructured mesh with boundary layer grids, (a) the mesh around the tube, (b) the mesh around the staggered bundle of tubes, (c) a close up of the mesh around the staggered tube banks.}
	\label{3-1}
\end{figure}

Overall and close snapshots of the computational mesh are presented in Fig.~\ref{3-1}. As it is seen, near the surface of tubes and fins, the structured boundary layer mesh is employed while an unstructured mesh is utilized for regions far from the surface. The size of the first cell on the body is very important while simulating the turbulent flow. In the near wall region, due to the intense gradient of parameters under the effect of viscosity, the center of the first cell of the mesh from the surface should be in a suitable distance from the wall. So, the non-dimensional number $y^+$ is defined as $\frac{yu_{\tau}}{\nu}$. The friction velocity is a non-dimensional quantity that is defined by the shear stress on the wall, the representative of the velocity gradient of the wall, and is explained as $u_{\tau}=\sqrt{\tau_{w}/\rho}$. The flow in the range 30$<y^+<$300 is completely turbulent and the Reynolds stress tensor overcomes the stresses resulted by the molecular viscosity. The center of the first cell should be placed inside the laminar sub-layer, to make the results to be reasonably accurate. In this study, the mesh is generated in a such way that the $y^+$ equals 1. The rate of growth in size of the boundary layer mesh is 1.1 to make sure that there are appropriate number of cells in the near-wall region.

\section{Validation of results}
\label{validation}
At the first step, the obtained results are compared to the reported data in previous researches in order to make sure that the present simulations present acceptable accuracy. Hence, initially, the mesh- and time-step size independence tests are performed, and then appropriate mesh/time-step sizes for the simulations are proposed. Secondly, the acquired results for heat transfer and flow around the fixed and  the oscillating cylinders will be compared with the available data.

\subsection{Mesh/time-step size independence tests}
Simulations have been performed on three different mesh topologies at Re=3900 to achieve the grid-independent results. The details of topologies and corresponding results are summarized in Tab. \ref{Tab:3-1}. When the number of elements of the computational grid increases from 29,516 elements to 39,516; the obtained results for the drag coefficient, the Nusselt number, and the Strouhal number, respectively, change for 0.18$\%$, 0.77$\%$, 0.42$\%$. Consequently, the grid with total size of 29,516 is adequate to accomplish an authentic simulation.

On the other hand, the numerical results were also investigated using three different time-step sizes. The Nusselt number, the drag coefficient and the Strouhal number are calculated then. As it is manifested in Tab. \ref{Tab:3-2}, the results for the normalized time-step of 0.001 is very near to that of 0.0001. In more details, the percentage change for the drag coefficient and the Nusselt number is less than 0.15\%, and less than 1.3\% for the Strouhal number. Accordingly, the case with 0.001 has been taken to proceed the calculation.

\begin{table}[htbp]
	\caption{The result of mesh independence test for the flow around two-dimensional fixed cylinder at Re=3900.}
	\label{Tab:3-1}
	\centering
	\begin{small}
		\hspace*{-0.9cm}
		\begin{tabular}{cccccc}
			\hline
			y$^+$  &  St  &
			$\bar{Nu}$  & C$_d$  & No. of cells  \\ \hline
			\ 0.99 \ & \ 0.236 \ & \ 44.718 \ & \ 1.628 \ & \ 17814 \ \\ \hline
			0.69 & \ 0.233 \ & \ 43.918 \ & \ 1.640 \ & \ 25242 \  \\ \hline
			0.68 & \ 0.236 \ & \ 43.730 \ & \ 1.631 \ & \ 29516 \  \\ \hline
			0.67 & \ 0.237 \ & \ 43.395 \ & \ 1.628 \ & \ 39578 \  \\ \hline
		\end{tabular}
	\end{small}
\end{table}
\begin{table}[htbp]
	\caption{The result of time-step size independence test for flow around two-dimensional fixed cylinder at Re=3900.}
	\label{Tab:3-2}
	\centering
	\begin{small}
		\hspace*{-0.9cm}
		\begin{tabular}{cccccc}
			\hline
			St  &  $\bar{Nu}$  &
			C$_d$  & Time step  \\ \hline
		\ \ \	0.254 \ \ \ & \ \ \ 43.772 \ \ \  &  \ \ \ 1.631  \ \ \ & \ \ \ 0.01  \ \ \ \\ \hline
		\ \	\ 0.236  \ \ \ & \ \ \ 43.730 \ \ \ &  \ \ \ 1.631  \ \ \ &  \ \ \ 0.001  \ \ \ \\ \hline
		\ \	\ 0.233 \ \ \ & \ \ \ 43.795 \ \ \ &  \ \ \ 1.629  \ \ \ &  \ \ \ 0.0001  \ \ \ \\ \hline
		\end{tabular}
	\end{small}
\end{table}

\subsection{Fixed cylinder}
In this section, validation of the present obtained results for the fixed cylinder are demonstrated. For this purpose, the results are validated comparing the calculated pressure coefficient with the data reported in~\cite{Rajani2016}. Also, the validation of the thermal part is performed via comparing the obtained local Nusselt number distribution with the available data reported in~\cite{Bouhairie2007}. In \cite{Rajani2016}, the flow around the cylinder in subcritical regime is simulated utilizing the large eddy simulation method. The results obtained for the pressure coefficient are, respectively, validated and verified with the reported findings in \cite{Rajani2016} and \cite{Norberg1987}.

\begin{figure}
	\centering
	\includegraphics[width=1.05\columnwidth]{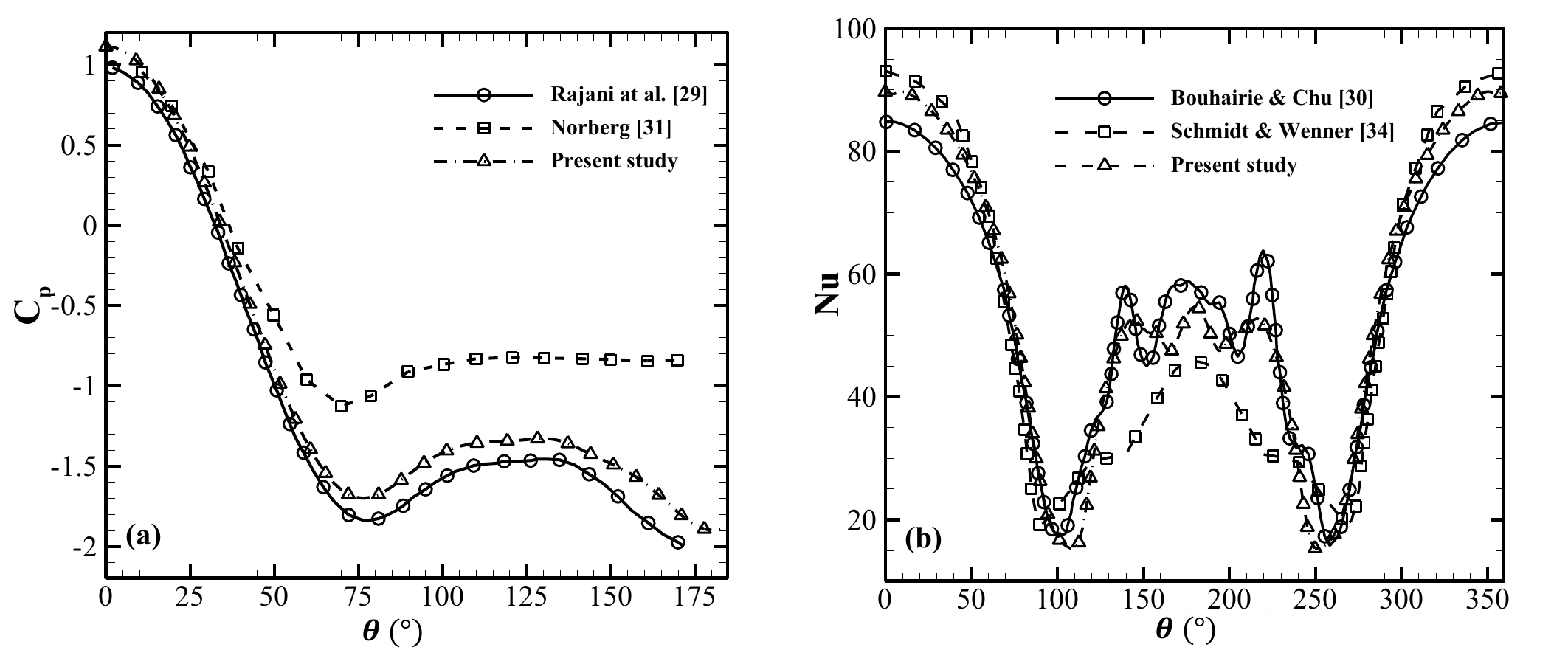}
	\caption{Validation of (a) the pressure coefficient distribution, (b) the local Nusselt number of the fixed cylinder in turbulent flow, when Re=8290.}
	\label{Fig3-2}
\end{figure}

Figure~\ref{Fig3-2} displays this specific comparison. As it is clear in Fig. \ref{Fig3-2} a, the difference between all three graphs is negligible at the separation point and it gets larger going farther from this point. At the angles close to the backward stagnation point, the calculated pressure coefficient in two-dimensional cases deviates from the experimentally obtained pressure coefficient presented in \cite{Norberg1987}. The numerical analysis predicts that the pressure coefficient decreases when getting closer to the backward stagnation point, while the experimental study presents a steady trend. This deviation has been also reported and more, justified by Mittal \cite{Mittal2001}. In experiments and the three-dimensional simulations, the cylinder side walls or say the end plates, make the formed vortices obliqued or in other words cause the formation of non-parallel vortices and also their exitance from the longitudinal direction of the cylinder, the phenomenon which do not occur in two-dimensional cases. Actually, in absence of curved vortices; the small-scaled effects, which only exist in three-dimension, still reduce the minimum values in the lift and drag oscillations. In three dimensional experiments and simulations, the presence of oscillation along the centerline of the cylinder and also the existence of vorticity component along the flow, results in notable difference in the Reynolds stress and all components of the stress in flow wake. This effect in turn, affects the pressure distribution, which itself impresses the drag force.


Furthermore, to confirm the accurateness of the calculations related to the heat transfer around the cylinder, the obtained local Nusselt number distribution about the cylinder has been compared with that of calculated in \cite{Bouhairie2007}. In this work, the heat transfer around the cylinder has been studied numerically. The local Nusselt number for Re=8290 has been compared with the experimental results of Schmidt and Wenner \cite{Schmidt1943}. In Fig. \ref{Fig3-2} b, these two previous findings as well as the present result for the local Nusselt number are shown. As Fig. \ref{Fig3-2} b confirms, the numerical results of the present study together with the numerical calculated Nusselt number of \cite{Bouhairie2007}, slightly differ from the experimental results reported in Wenner \cite{Schmidt1943} just after the separation point and in the positions near the rear side stagnation point of the cylinder. This discrepancy is attributed to the presence of the three-dimensional effects in experiments. This difference is much less recognized in positions placed before the separation point. More particularly, near the frontal stagnation point, the obtained results are close to the experimental data.

\subsection{Oscillating cylinder}
\begin{figure}
	\centering
	\includegraphics[width=0.8\columnwidth]{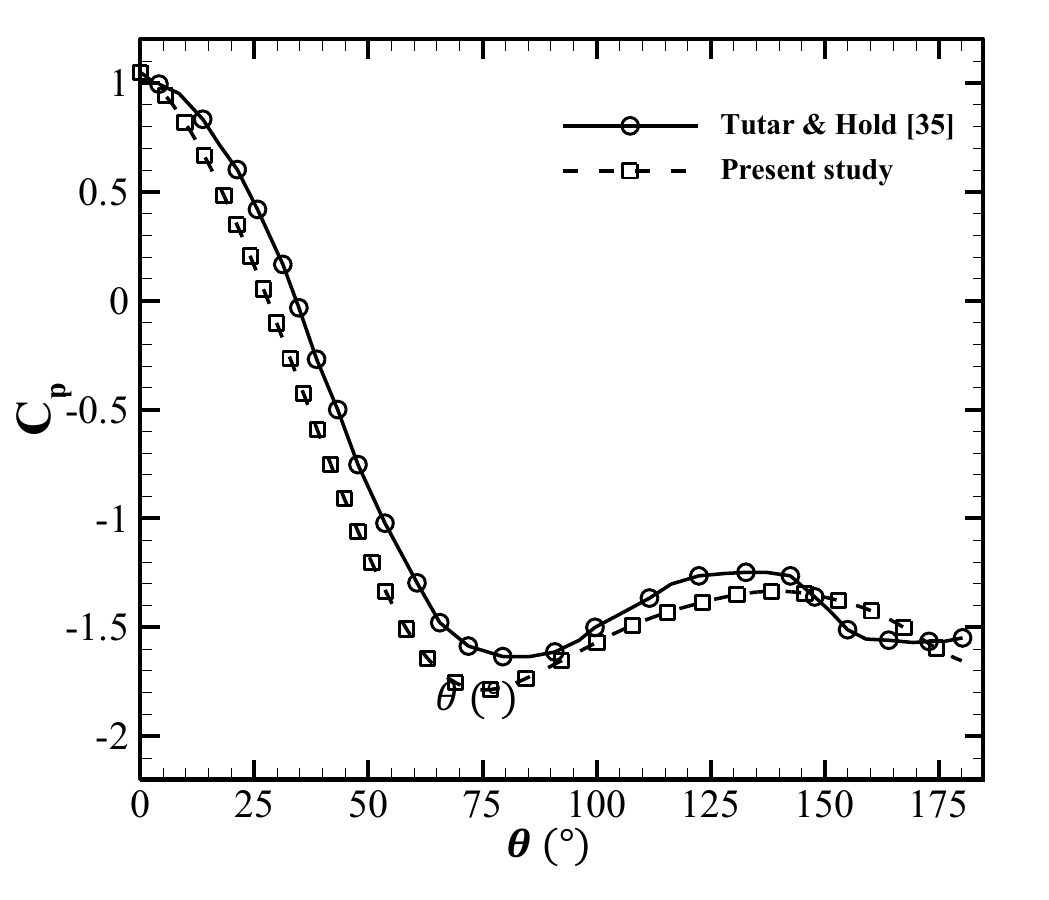}
	\caption{Validation of the pressure coefficient distribution around oscillating cylinder in turbulent flow at Re=24,000.}
	\label{Fig3-3}
\end{figure}

In this section, the obtained results for the pressure coefficient and the local Nusselt number are presented and compared with previous data for oscillating cases. The pressure coefficient is validated using the available data in \cite{Tutar2000}, where the big vortices around an oscillating cylinder are numerically investigated in two and three dimensions. Fig. \ref{Fig3-3} demonstrates the available data for pressure coefficient distribution for two-dimensional cases at Re=24000 \cite{Tutar2000} and the present acquired pressure coefficient. As it is obvious, the error between two plots is about 0.2 before and after the separation point.

\begin{figure}
	\centering
	\includegraphics[width=0.8\columnwidth]{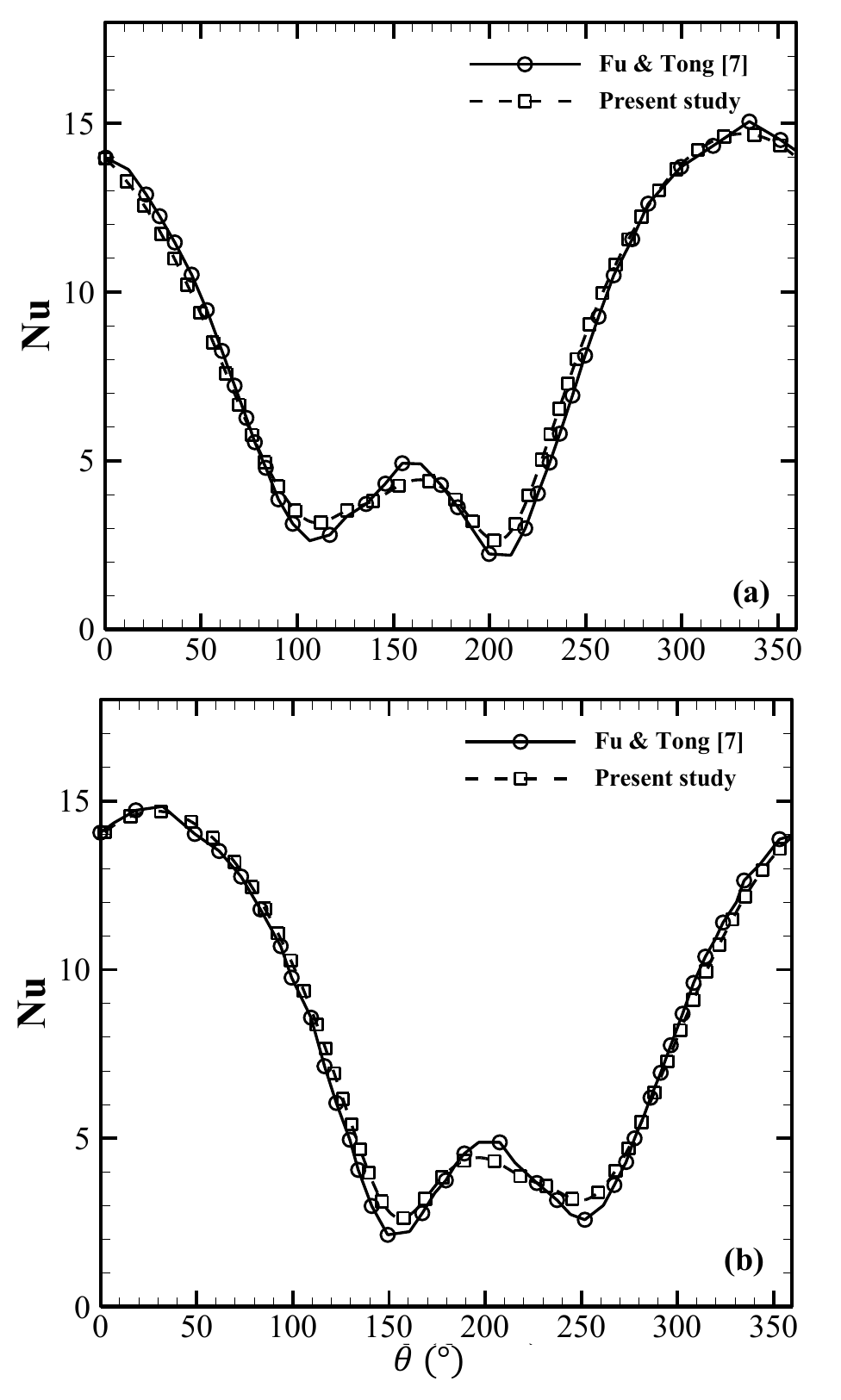}
	\caption{Validation of the local Nusselt number distribution around the oscillating cylinder in Re=200 for (a) t=T/4, (b) t=3T/4.}
	\label{Fig3-55}
\end{figure}

In order to check the accuracy of the dynamic mesh part of the solver, the obtained local Nusselt number distribution around an oscillating cylinder has been compared with the results calculated in \cite{Fu2000}. In this study, the flow structure and the heat transfer about the oscillating cylinder is  investigated where the effects of the Reynolds number, and the amplitude have also been inquired. The available reported data and also our obtained results at four different times during one period of oscillation at Re=200, A/D=0.4, $f^*$=0.2, are displayed in Fig.~\ref{Fig3-55}. At t=3T/4 and t=T/4, the cylinder is, respectively, at its upper and lower positions during a cycle. Consequently, the difference between the plots of the Nusselt number distribution is increased up to 8$\%$, while this difference for t=T and t=T/2 when the cylinder is placed in the origin of oscillating is less than 0.2$\%$.

\section{Results and discussion}
\label{results}
Table~\ref{Tab:4-11} presents list of details of different cases studied in this section. Input parameters include position of slot, breadth of slots, number of slots, amplitude of oscillation, height of fins, column of oscillatory excitation, being slotted, inserting fins, single/bank of tubes.

\begin{table*}[htbp]
\caption{Different case studies and the value of related variables.}
	\centering
	\begin{small}
		\hspace*{-0.5cm}
		\begin{tabular}{cccccccccc}
			\hline
			Cases  &  Fin  &
			Slot  & Oscillation  & \makecell{Slot \\ position (r$^*$)} & \makecell{Slot \\ breadth (e$^*$)} & \makecell{No. of \\ slots (N)} & \makecell{Amp. of \\ Osc. (A$^*$)} & \makecell{Height of \\ Fin (H)} & \makecell{Oscillating \\ column} \\ \hline
			\makecell{Fixed cylinder \\ without fin} & \ & \ & \ & - & - & -& - & - & -  \\ \hline
			\makecell{Fixed cylinder with \\ fin without slot} & \checkmark & \ & \ & - & - & - & - & 0.75 & -  \\ \hline
			\makecell{Fixed cylinder with \\ slotted fin} & \checkmark & \checkmark & \ & \makecell{0.05-0.15 \\ 0.25} & \makecell{0.01-0.02-0.04\\ 0.06} & 1-2-3 & - & \makecell{0.65-0.75 \\ 0.85} & -  \\ \hline
			\makecell{Oscillating cylinder \\ with slotted fin} & \checkmark & \checkmark &  \checkmark & \makecell{0.05-0.15 \\ 0.25} & \makecell{ 0.02-.0.04} & 3 & 0.2-0.4-0.6 & 0.65 & - \\ \hline
				\makecell{Fixed bundle of \\ tube without fin} & \ & \ & \ & - & - &- & - & - & - \\ \hline
			\makecell{Oscillating bundle \\ of tube without fin} & \ & \ & \checkmark & - & - & - & 0.4 & - & 1-2-3-4  \\ \hline
				\makecell{Oscillating bundle of \\ tube with slotted fin} & \checkmark & \checkmark & \checkmark & \makecell{0.05-0.15 \\ 0.25} & 0.02-0.04 & 3 & 0.4 & 0.65 & 1-2-3-4\\ \hline
		\end{tabular}
	\end{small}
	\label{Tab:4-11}
\end{table*}

\subsection{Fixed finned cylinder}
\label{Case1}
\begin{figure}
	\centering
	\includegraphics[width=\columnwidth]{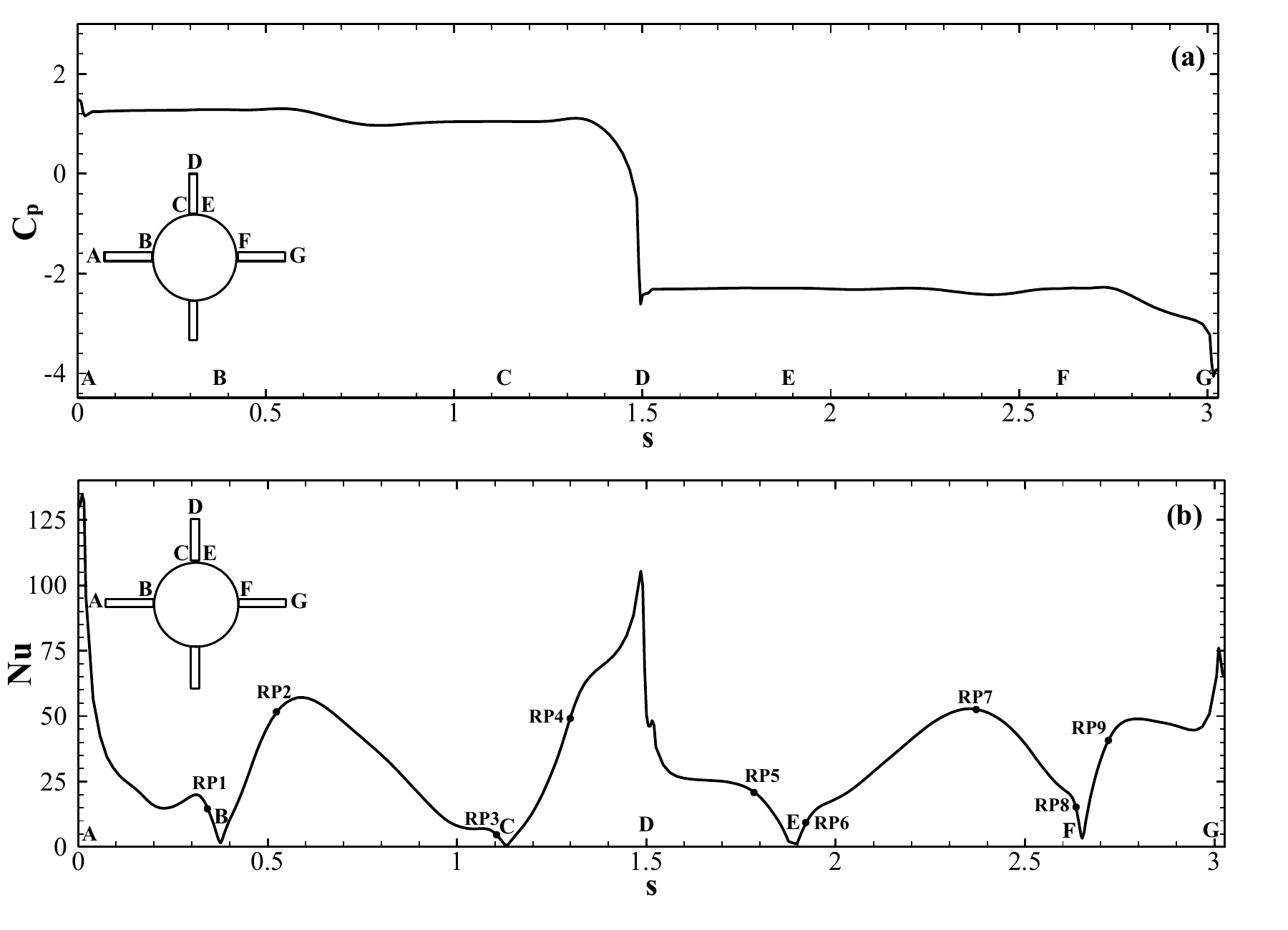}
	\caption{(a) The pressure coefficient distribution and (b) the local Nusselt number around the finned cylinder without slot when H=0.75.}
	\label{Fig4-322}
\end{figure}

Figs. \ref{Fig4-322} a and b demonstrate the pressure coefficient and the local Nusselt number distributions over the finned cylinder without slot, for different "s" parameter, which is the coordinate attached to the surface of the cylinder. The coordinate "s" starts from s=0 located at the tip of the fin number 1, and continues up to s=3.02 at the tip of the fin number 4. s$<$1.5 corresponds to the low-pressure region, or say the region in the right-hand side of the fin number 2.

In order to better discuss the pressure coefficient distribution and the Nusselt number diagrams, points A to G are considered at the upper half of the finned tube, and their corresponding points with the same names specified  at the "s"-axis. In more details, point A is placed at the tip of the fin$\#$1, which is displayed at position s=0. B denotes fin$\#$2 connection to the surface of the cylinder, and corresponds to s=0.37. Also, points C and E are the left and right roots of the fin number 2 respectively, and are shown at s=1.13 and s=1.18. The D point is located at the tip of the fin$\#$2 and divides the plot into two parts. Points F and G also correspond to fin$\#$4 and are placed, respectively, at the jointing points of the fin to the cylinder and the tip.

As it is obvious in Fig. \ref{Fig4-322}, the points being implemented subsequently at s=2.65 and s=3.02, are presented in both the "s"-axis and the schematic plot. The time-averaged pressure for all the points with s<1.5, except the tip of fin$\#$1 where the pressure is slightly higher, is approximately uniform. The pressure coefficient in the left region of the fin$\#$2, from point A to D is positive, while it is uniformly negative in the right side of the point D. It takes the lowest value in the point close to the tip of the fin$\#$4, or say G. The pressure coefficient suffers a sharp reduction from the positive value to the negative one at the tip of the fin$\#$2. The pressure difference between the two sides determines the value of the pressure (form) drag coefficient. As much as the pressure difference grows, the pressure drag coefficient gets larger. Moreover, the drag coefficient of the cylinder with four fins, C$_d$=3.369, is found to be almost two times larger than that of the cylinder without fin, C$_d$=1.631.

In Fig. \ref{Fig4-322} b, the Nusselt number distribution versus "s" has been illustrate. The maximum value of the Nusselt number occurs for the stagnation point A, placed at the tip of the fin$\#$1. The value of Nu decreases going from point A to the point B. In vicinity of the joint point of the fin to the cylinder, around point B, a relative maximum of the Nusselt number is conspicuous. The points denoted by RP$_i$ with i being 1 to 9 and marked in Fig. \ref{Fig4-322} b, are the points for which in closeness, the relative maximum of the Nu occurs. Each of these points, is representative of the reattachment point that is the point where the dividing streamline reattaches to the cylinder wall. Also, Zdanski \emph{et al.} \cite{Zdanski2016} mentioned this Nu augmentation near the reattachment points, studying the backward facing step case with considering a turbulence promoter. This Nu increase is attributed to the heat transfer rate augmentation caused by the increment of the turbulence diffusion mechanism.

\begin{figure}
	\centering
	\includegraphics[width=0.9\columnwidth]{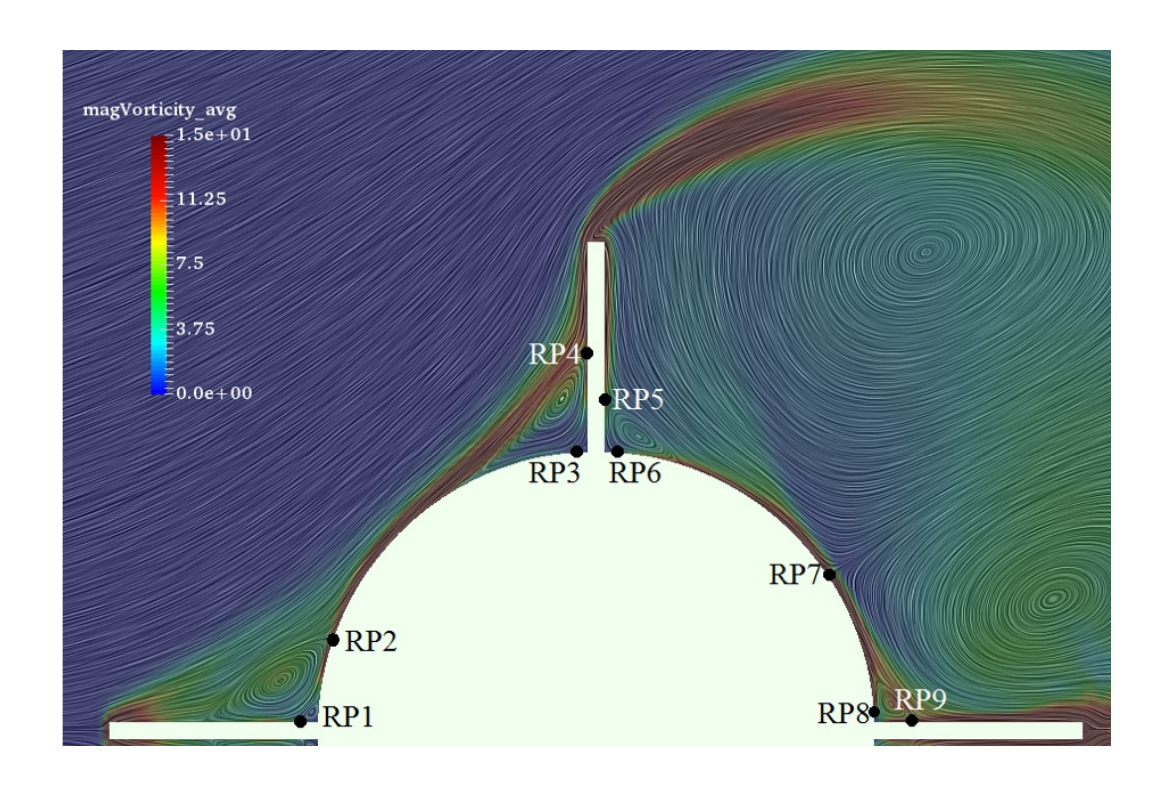}
	\caption{Distribution of reattachment points in time-averaged vorticity field around the finned cylinder without slots.}
\label{Fig4-44}
\end{figure}

The time-averaged vorticity field is presented in Fig.~\ref{Fig4-44}. The points denoted as RP can be also seen in Fig.~\ref{Fig4-322} b. As it is recognized, the flow detaches from the surface at the sharp tip of the fin number one, reattaches to the cylinder at point RP2. Fluid flows over the surface, and again separates from the cylinder surface. The separated flow at point RP4, attaches to the fin$\#$2. In the right half of the body, the separated flow forms a vortex at the sharp tip of the fin$\#$2. The sharp tip of the fin$\#$4 creates another bubble. The separated flow at the tip of the fin reattaches to the cylinder surface at point RP7. Similar phenomenon takes place in the neighborhood of points E and F in smaller scale. In vicinity of the mentioned points, a local maximum for the Nusselt number is detected.

The local Nusselt number distribution presented in Fig. \ref{Fig4-322} is similar to the one reported in \cite{Abu-Hijleh2003}. In the mentioned study, the fins are considered thicknessless. The Nu distribution is presented versus the angle on the surface of the cylinder. As results in \cite{Abu-Hijleh2003,Bouzari2016} suggest, the Nu on the cylinder surface increases from the connection point of the fin to the cylinder. Reaching the relative maximum, it obeys the reduction trend until it catches the connection point of the next fin to the cylinder. Just before point C, a relative maximum appears in Nu distribution owing to the existence of the point RP3. The Nu on the surface of the Fin$\#$2, increases from the point C that takes its highest value in the tip of the Fin.

In vicinity of the fourth reattachment point, a secondary extremum is apparent in the Nu distribution. In tip of the Fin$\#$2, due to the compression of streamlines, the fluid takes a high velocity and the Nu increases. In other side of this fin, from the tip of the fin to its connection point to the cylinder, the Nu follows a decreasing trend, and due to the RP5 existence, a relative maximum of the Nu is conspicuous. On the surface of the cylinder, between fins$\#$2 and $\#$4 or say from point E to the point F, there exist three reattachment points in the vicinity of which the relative maximums are recognized. Generally, between the points E to F, first the Nu increases and then it decreases until the root of the Fin$\#$4. Also, the growing trend of the Nu is evident over the Fin$\#$4. The convection heat transfer coefficient augments with increasing the vorticity due to the higher mixing of the swirling flow.

\begin{figure}
	\centering
	\includegraphics[width=1.1\columnwidth]{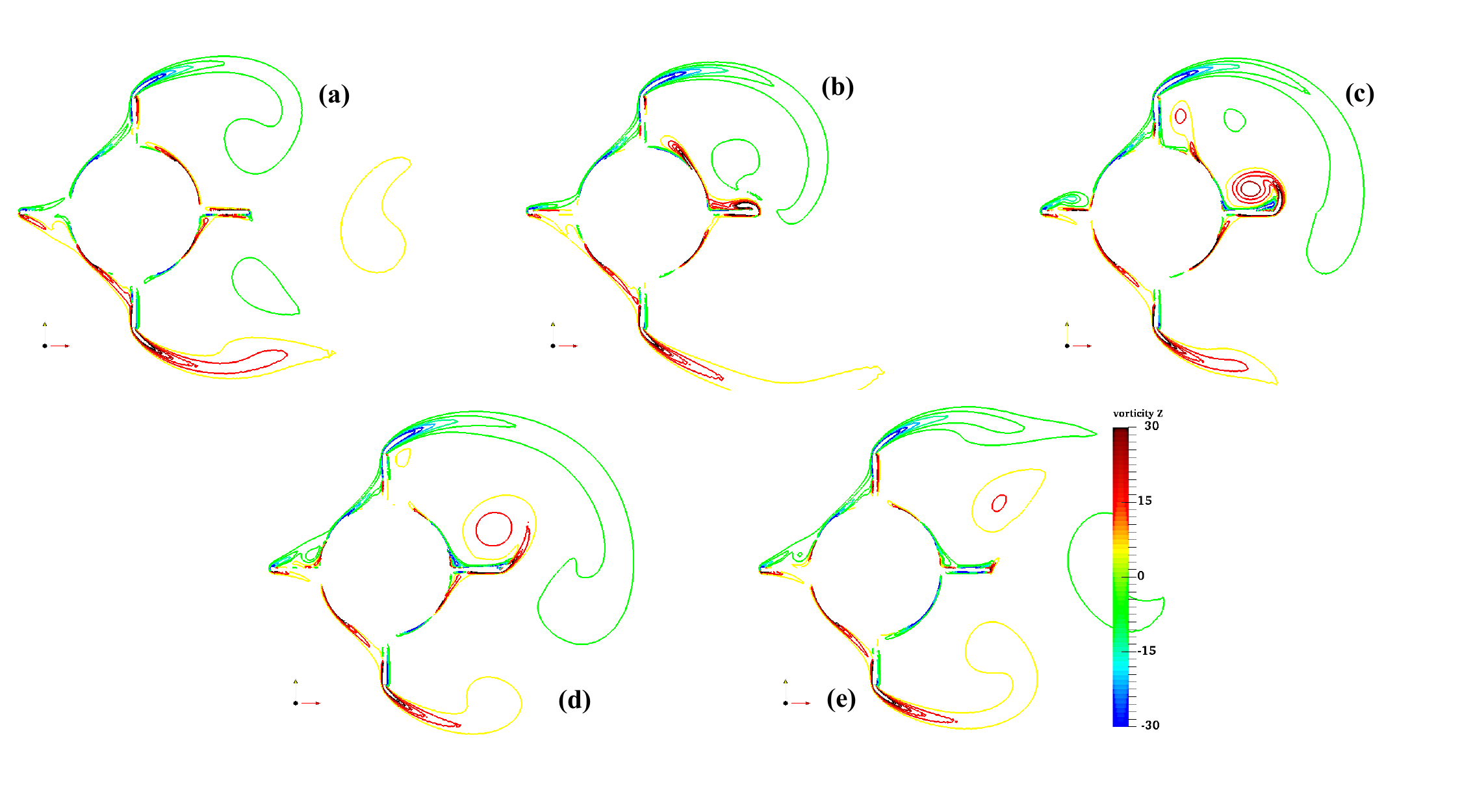}
	\caption{The vorticity contour around the finned cylinder without slot during the half period when (a) t=0, (b) t=T/8, (c) t=T/4, (d) t=3T/8, and (e) t=T/2.}
	\label{Fig4-50}
\end{figure}

The vorticity contour around the finned cylinder without slot during the half period, has been presented in Fig. \ref{Fig4-50}. In the second half period, a similar happening occurs in the lower half of the body. As it is recognizable, the flow separates at the tip of fin number one. At this point the value of the vorticity is high. The vortex resulted from this separation, affects the Nusselt number on the surface of the Fin$\#$1. Specially near the reattachment point that is close to the point B, the Nu increase is evident. After flowing the fluid on the surface, it separates from, and subsequently attaches to the fin number two. This leads to the formation of a region with low vorticity in vicinage of the point C and Nu decreases in this zone.

In the right side of the fin number two, the flow structure is more complex. Several vortices are created in the region between the fins$\#$2 and $\#$4, which affect the flow field and consequently the heat transfer. The flow separates at the sharp tip of the fin$\#$2 and a vortex is formed in the downstream. As Fig. \ref{Fig4-50} b suggests, this separated vortical flow hits the fin number 4, and owing to the existence of the sharp point on the tip of this Fin, separates from it and another vortex is created. Indeed, Fig. \ref{Fig4-50} b presents the creation of another vortex simultaneously with this vortex, on the cylinder surface. The two vertices grow together and each one affects the thermal condition on a part of the surface of the body. Figure \ref{Fig4-50} c presents the growth of these two vertices. The formed vortex at the tip of the fin$\#$4 influences the flow pattern over the fin and the cylinder surfaces in the vicinity of point F.


\subsection{Fixed slotted finned cylinder}
\label{Case2}
As mentioned in the previous section, adding fins to the cylinder makes the drag coefficient to increase and the Nusselt number to decrease. In this section, to equalize the pressure distribution around the fin, slots on the vertical fins will be considered in which the flow can pass through. These slots modify the drag coefficient by creating the connection between the high-pressure and low-pressure sides. Also, the fluid flow passing through these slots changes the flow structure in the area between the fins$\#$2 and $\#$4. Effects of the parameters such as, the position of slot placement or its distance from the surface (r$^*$), the width of the slot (e$^*$), the number of slots on the fin (N), the length of the slot (H), and the oscillation amplitude of the cylinder (A$^*$) on the drag coefficient, the Nusselt number, the fin effectiveness, and the Nu-to-drag coefficient ratio will be discussed.

\subsubsection{\label{2.4.1}Position of the slot}
\begin{figure}
	\centering
	\includegraphics[width=\columnwidth]{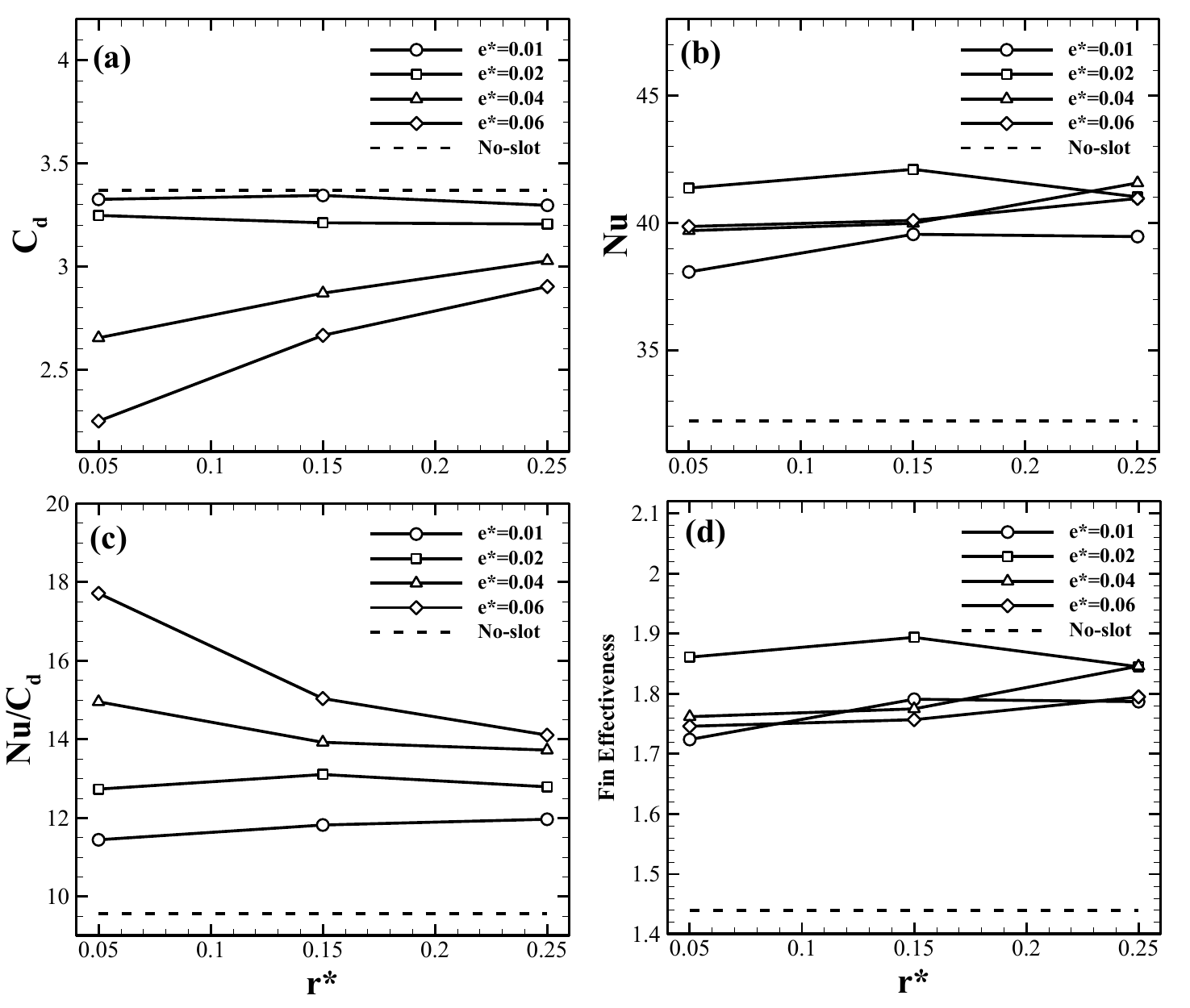}
	\caption{(a) The drag coefficient, (b) the Nusselt number, (c) the Nusselt Number to the drag coefficient ratio, (d) the fin effectiveness versus r$^*$ for case with different slot widths, e$^*$.}
	\label{Fig4-7}
\end{figure}

Three different dimensionless distances of r$^*$=0.05, r$^*$=0.15, and r$^*$=0.25 for the slot on the fin are considered. Also, the slot with four different widths of e$^*$=0.01, e$^*$=0.02, e$^*$=0.04, and e$^*$=0.06 is located individually in the mentioned positions, r$^*$. Fig. The drag coefficient versus r$^*$, has been plotted in Fig.\ref{Fig4-7} a. As it is obvious, in all cases of the cylinder with the slotted fin, the drag coefficient is less than that of the one with a fin without slot. For e$^*$=0.01, the drag coefficient does not significantly differ relative to its value for the finned cylinder without slot. In this case, the drag coefficient remains almost constant by increasing the r$^*$ and it does not matter, how far the slot is from the cylinder. Also, when e$^*$=0.02, a slight reductive trend relative to the r$^*$ increment is seen. On the other hand, the drag coefficient enhances when r$^*$ increases, for e$^*$=0.04 and e$^*$=0.06. The drag coefficient presents maximum reduction of 33$\%$ relative to its value for the finned cylinder without slot. This maximum reduction occurs, for a slot with e$^*$=0.06, placed in the nearest distance from the cylinder, r$^*$=0.05. In general, when the slot is thin, the drag coefficient variation is less than 1.5$\%$, while for wider slots, this reduction is more prominent. Getting closer to the cylinder surface, the drag coefficient decreases more.

Fig. \ref{Fig4-7} b demonstrates the Nu variation versus r$^*$ for different values of e$^*$. For all values of r$^*$, placing slot on the fin, results in augmentation of Nu in comparison to the finned cylinder without slot. The behavior of the Nu versus r$^*$ for the two slots with smaller width of e$^*$=0.01 and e$^*$=0.02, is the same, such that firstly it increases and then decreases. For two other wider slots, the Nu continuously increases moving away from the cylinder surface. The maximum Nu, which is 30.7$\%$ larger, belongs to the slot with e$^*$=0.02 placed at r$^*$=0.15, or say in mid areas of the fin. 
The plot presenting the Nu/$C_d$ ratio is shown in Fig. \ref{Fig4-7} c. For all cases, an increase in value of this ratio for slotted finned cylinders is obvious. When e$^*$=0.01, due to the smooth variation of the C$_d$ and Nu versus r$^*$, Nu/C$_d$ ratio increases slowly. Owing to the slighter change of the drag coefficient, the ratio Nu/$C_d$ for e$^*$=0.02 obeys the same trend of the Nu, that is, augmenting at first and secondly reducing. For slots with e$^*$=0.04 and e$^*$=0.06, both Nu and C$_d$ increase while r$^*$ enlarges. As the drag coefficient change is much stronger than the Nusselt number variation, the ratio of these two dimensionless numbers decreases with r$^*$ increasing. The largest value of the Nu/C$_d$, being 85$\%$ higher than the value for finned cylinder without slot, corresponds to the widest slot placed in the closest distance from the cylinder surface. 

The fin effectiveness, which is defined as the ratio of heat transfer from the finned cylinder to the heat transfer through the cylinder without fin, versus r$^*$ is indicated in Fig. \ref{Fig4-7}. The Fin effectiveness is a function of Nu and the heat exchange surface area. By keeping e$^*$ constant, the heat exchange surface does not change as the slot moves away from the cylinder surface. Consequently, the fin effectiveness follows a trend similar to the Nu changes, for each e$^*$. The maximum value of the fin effectiveness for r$^*$=0.05 and r$^*$=0.15 belongs to the slot with e$^*$=0.02, and also, in the furthest distance from the cylinder surface, the highest fin effectiveness happens for e$^*$=0.04. In best situation, at r$^*$=0.15, a slot with the width of e$^*$= 0.02, whose effectiveness equals 1.894, increases the effectiveness of the system compared to the non-slot mode by 31.5$\%$.

\begin{figure}
	\centering
	\includegraphics[width=\columnwidth]{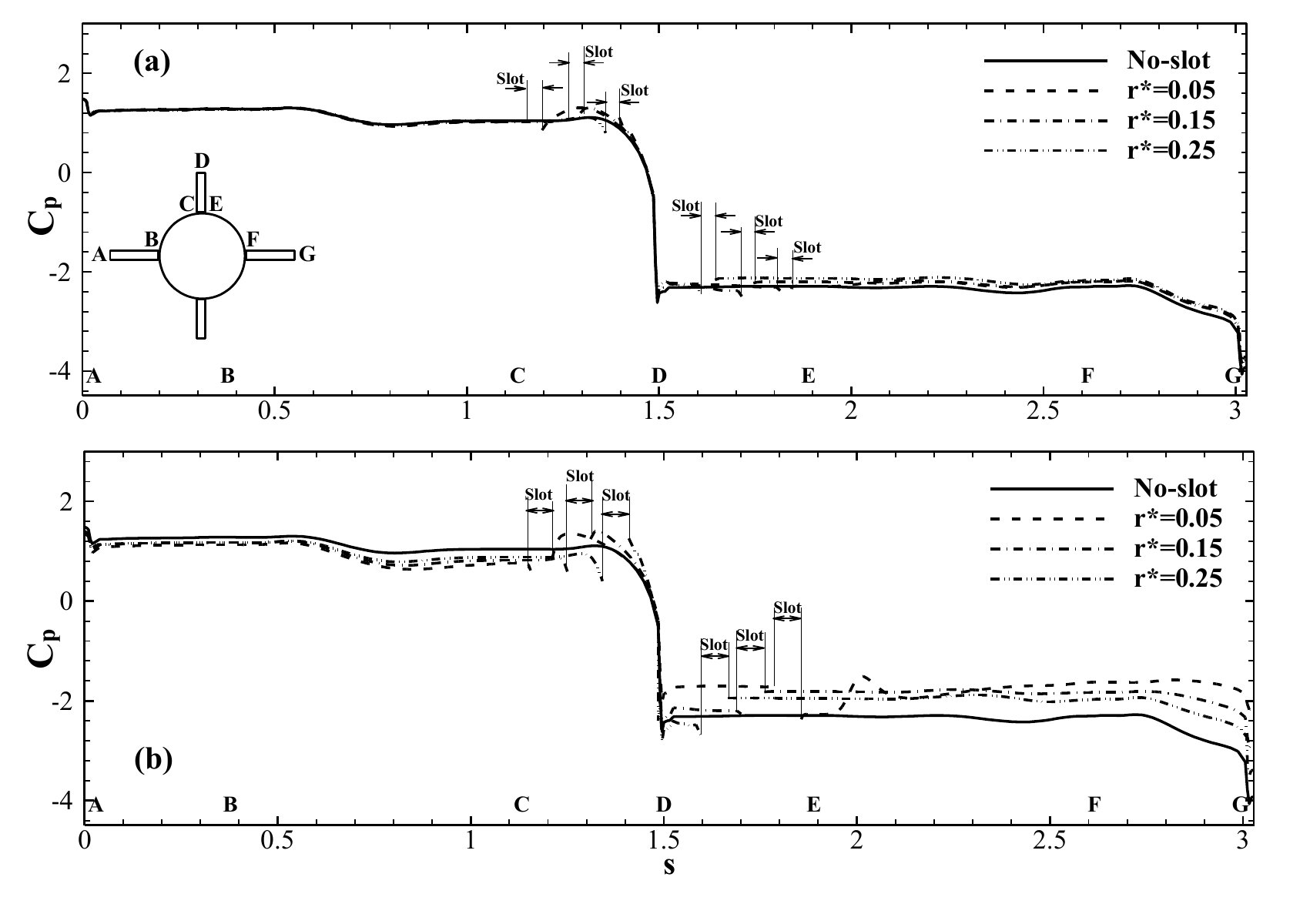}
	\caption{The pressure coefficient distribution of the cylinder with slotted fin configuration for different values of the slot distances from the surface of the cylinder with slot thickness of: (a) e$^*$=0.01, (b) e$^*$=0.02.}
	\label{Fig4-8}
\end{figure}

The drag coefficient variations result from the change of pressure coefficient on the surface of the body. In order to investigate origins of the behavior of drag coefficient variation with r$^*$, the distribution of the pressure coefficient on the finned cylinder with slots, for different values of e$^*$ is presented in Fig. \ref{Fig4-8}. Fig. \ref{Fig4-8}(a) shows the schematic of the cylinder with slotted fins and the letters A to G on it in the insert. Also, on the "s" axis, the letters from A to G are seen, which are the points corresponding to the schematic shape on the surface of the object. The point D placed in the tip of the fin$\#$2, divides the plot into two halves. The position of the slots can be seen in both sides of the point D on the fin number 2. The points C to D present the left side of the fin$\#$2, where the slot in this area, indicates the notch of fin surface on its left side. On the other hand, points D to E is related to the right side of the fin and the presented slot shows the notch of the right-side of the fin surface caused by it. 

Fig. \ref{Fig4-8} also demonstrates the effect of the distance of the slot from the cylinder surface, for slots with different widths. As it is obvious, placing the slot somewhere from point C to D on the fin, leads to the increasement of pressure coefficient in the left side of the part of the fin above the slot. This is expected since this part of the fin acts like a separating object in the flow path and a high-pressure stagnation point is formed on its front. Due to vortices formed behind this area, a low-pressure area is created in the backside.

\begin{figure*}
	\centering
	\includegraphics[width=1.8\columnwidth]{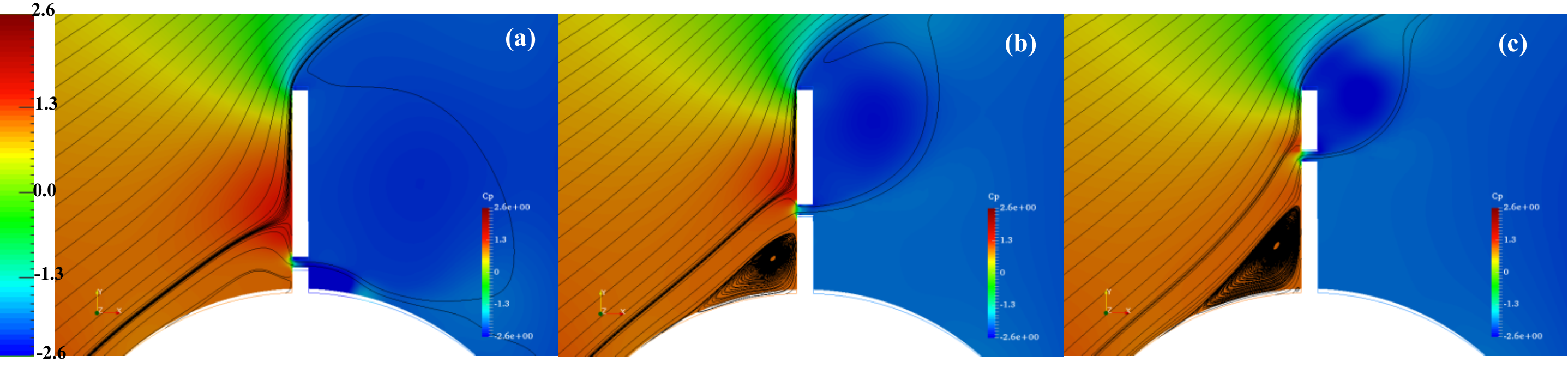}
	\caption{Formation of the stagnation point and the coherent vortex structure near the above zone of the slot with e$^*$=0.02 for distances from the surface equal to (a) r$^*$=0.05, (b) r$^*$=0.15, (c) r$*$=0.25.}
	\label{Fig4-9}
\end{figure*}

Figure. \ref{Fig4-9} demonstrates the created stagnation point in the mentioned zone. As the slot moves away from the cylinder surface, the high-pressure area becomes smaller and its pressure decreases. It is also seen in Fig.~\ref{Fig4-8} that the farther the slot is from the point D and the closer it is to the points C and E, the high-pressure area in front with higher pressure is wider. Furthermore, as the slot moves away from the cylinder surface, the size of the vortical area  in the wake becomes smaller and the low-pressure area becomes closer to the body surface, and consequently reduces the pressure on the right-side of fin surface above the slot. At the lower part of the slot, the trend is inverse of that of the above one. In other words, the closer the slot is to the cylinder surface, a smaller and closer vortex is formed that results in reduction of the pressure near this part.

Comparison of Fig. \ref{Fig4-9} a and Fig. \ref{Fig4-9} c presents, as the slot gets closer to the cylinder surface, a more distinct vortex eye in the backward of the lower part of the fin. On the other hand, when the slot is away from the cylinder and closer to the tip of the fin, an evident vortex eye is apperceived at the behind of the top of the fin. Due to the low pressure of the vortex eye, as the slot moves away from the cylinder, the low-pressure vortex eye behind the lower part of the slot moves away from the surface, and the vortex center approaches the surface at the top of the slot.

Shrinkage of the high-pressure region above the slot on its left side and increasing the pressure of the lower right side of the fin reduces the drag coefficient. Fig. \ref{Fig4-8} a is correspondent to the slot with e$^*=0.01$, which has no effective effect on the drag coefficient distribution at any location, r$^*$, it has been placed. In this case, the crossing fluid flow through the slot is weak, has a low velocity and penetration capability, and can not affect the wake behind the fin. Hence, the drag coefficient changes are low and almost constant for all r$^*$. The drag coefficient follows the pressure distribution and although it behaves nearly unchanging, but still its small maximum occurs in the middle areas of the fin arounf r$^*$=0.15. 

Fig. \ref{Fig4-8} b presents the results with e$^*$=0.02. In this case, the drag coefficient is also affected by the pressure distribution on the fin surface, except that at r$^*$=0.05, the pressure below the slot at right side of the fin$\#$2, decreases more in comparison to the one with e$^*$=0.01. This results in augmentation of the drag coefficient compared to the two other r$^*$. Here, also, the percentage of drag coefficient changes with r$^*$ is not high and is below 1.5$\%$. For two wider slots, the trend of drag coefficient changes with r$^*$ increasing, is incremental. For e$^*$=0.04 and e$^*$=0.06, the passing flow through the slot, changes the pressure distribution over the cylinder and the fin surfaces, as the jet crossing the slot is stronger and has a higher penetration capability than the narrower slots. In this matter, not only the passing flow rate through the slot is increased, but also, the crossing flow from the slot has higher momentum. Hence, it affects the wake of the body and also the formed vortices on its behind.

\begin{figure*}
	\centering
	\includegraphics[width=1.8\columnwidth]{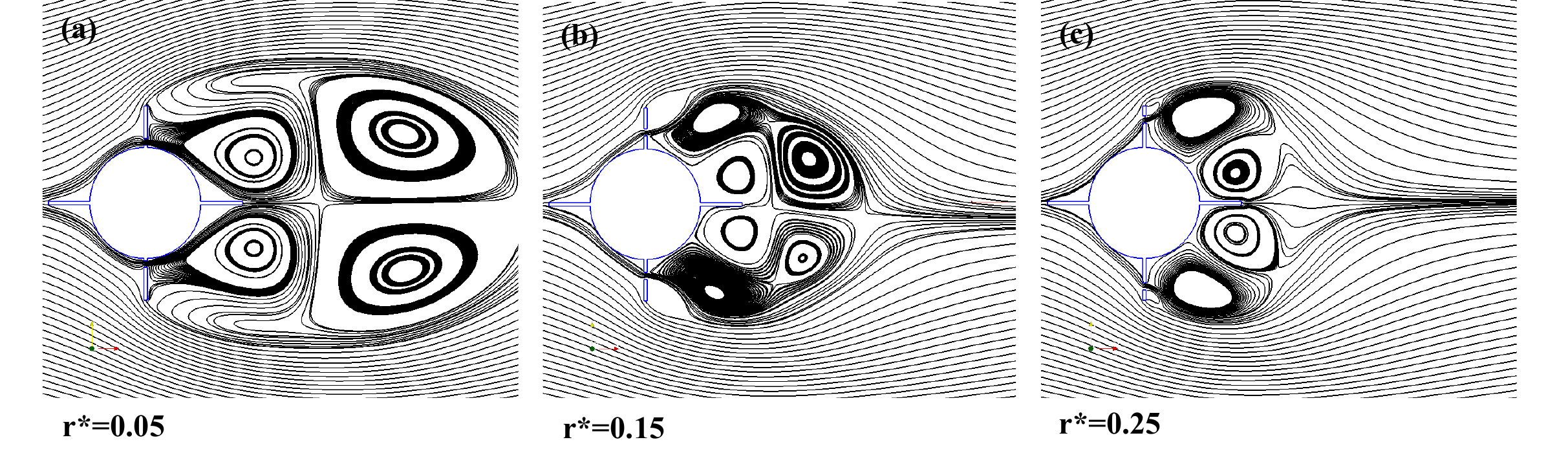}
	\caption{The time-averaged variation of the streamlines around the cylinder with slotted fin for e$^*$=0.06 with different values of slot distance from the fin root, (a) r$^*$=0.05, (b) r$^*$=0.15, (c) r$^*$=0.25.}
	\label{fig4-10}
\end{figure*}

Fig. \ref{fig4-10} acts to guide our discussion by plotting the time-averaged streamlines around the cylinder with different positions of the slots. In this case, due to the small flow rate of the fluid passing through the slot, the formed jet has lower momentum, and therefore has low penetration capability. Figs. \ref{fig4-10} a-c illustrating the streamlines when e$^*$=0.06 confirm that the closer the slot is to the cylinder surface, the larger the size of vortices becomes. As the length of the vortex formation area increases, and the low-pressure zone moves away, the surface pressure augments, which itself results in drag coefficient reduction. 

\begin{figure}
	\centering
	\includegraphics[width=\columnwidth]{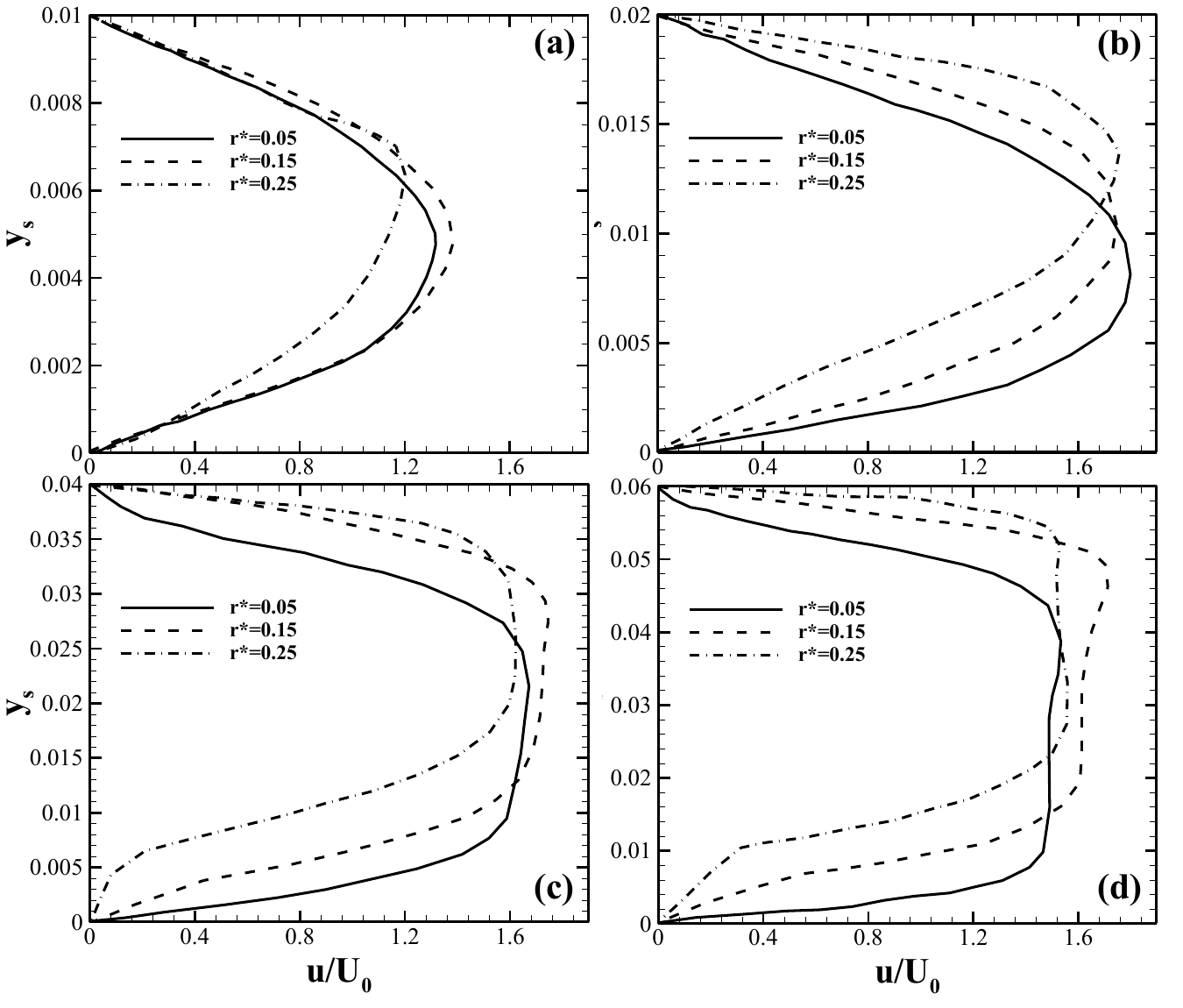}
	\caption{The velocity profile of the jet passing through the slot for different values of r$^*$, for (a) e$^*$=0.01, (b) e$^*$=0.02, (c) e$^*$=0.04, (d) e$^*$=0.06.}
\label{Fig4-11}
\end{figure}

The Jet velocity profile passing through the slot is presented in Fig. \ref{Fig4-11}. According to this figure, as the slot gets wider, the velocity profile tends from a parabolic state to a uniform profile, which marks the advent of turbulent flow. This slot broadening enlarges the passing jet area and the flow rate increases, as well, the passing jet becomes more turbulent with higher momentum to  penetrate. Consequently, the passing jet affects the pressure distribution over a broader part of the surface. The R-$\omega$ model in contrast to the large eddy simulation models, which has high power in recognizing the local laminarization of the flow, are not able to detect local laminar flow. Therefore, it is better to review the value of the turbulence viscosity in order to check the accuracy of the results obtained for the jet velocity profile passing through the slot for e$^*$=0.01 and e$^*$=0.02, where the parabolic velocity profile is obtained.

\begin{figure}
	\centering
	\includegraphics[width=\columnwidth]{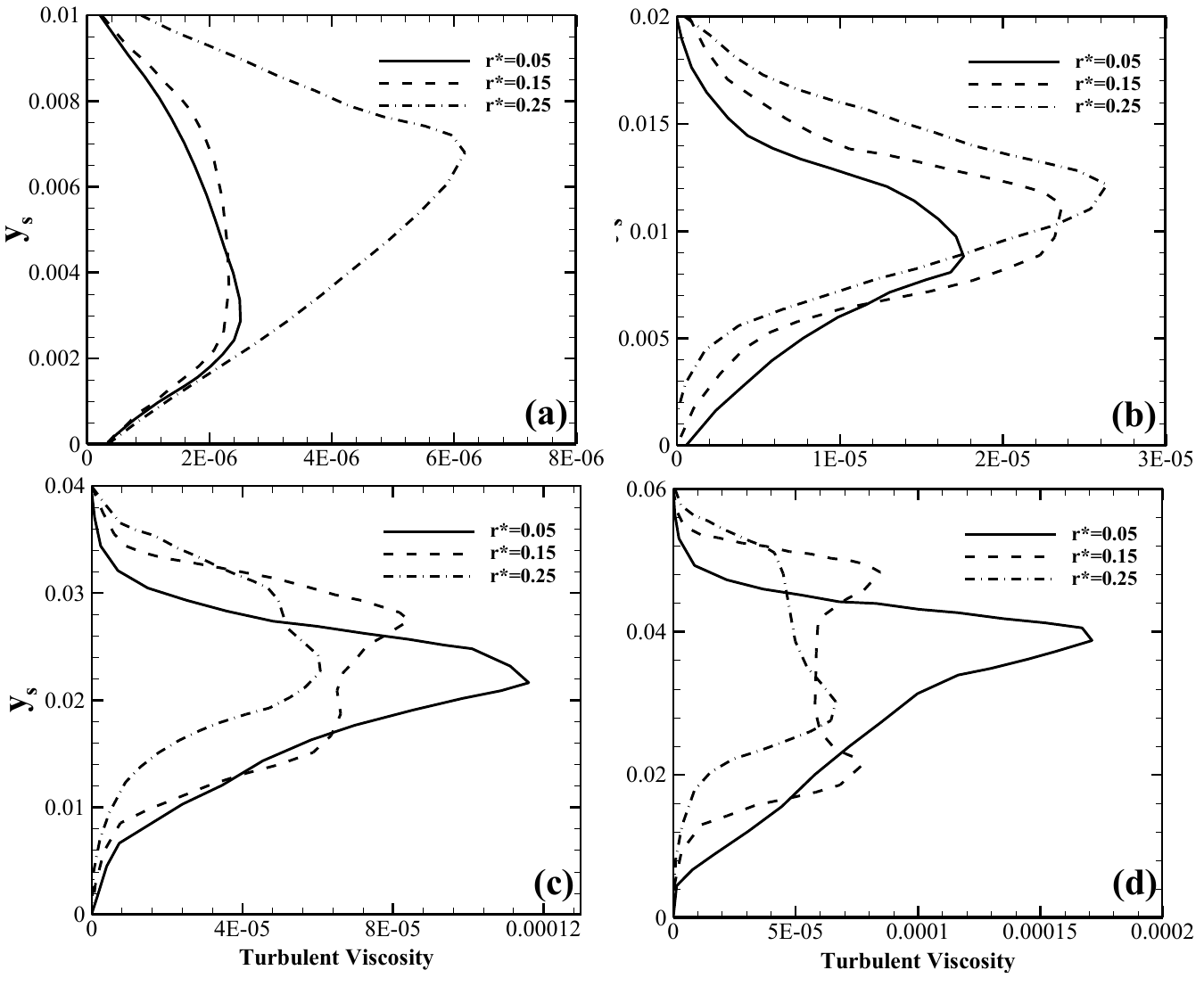}
	\caption{Distribution of the turbulent viscosity of the jet passing through the slot with different r$^*$ and different thicknesses: (a) e$^*$=0.01, (b) e$^*$=0.02, (c) e$^*$=0.04, and (d) e$^*$=0.06.}
	\label{Fig4-12}
\end{figure}

The turbulent viscosity of the passing jet for different values of e$^*$ placed in various positions is presented in Fig. \ref{Fig4-12}. As Figs. \ref{Fig4-12}(a) and (b) being respectively for e$^*=0.01$ and e$^*=0.02$ suggest, despite the parabolic velocity profile, which is the indicator of laminar flow, the turbulence viscosity is not exactly zero. However, the turbulence viscosity is very low for these cases, namely, being in order of magnitudes of $10^{-6}$ and $10^{^-5}$, accordingly, for e$^*=0.01$ and e$^*=0.02$. Also, the turbulence viscosity for e$^*=0.04$ and e$^*=0.06$, which is respectively shown in Figs. \ref{Fig4-12}(c) and (d), has higher value in comparison to the ones with narrower slots. This increasement shows the level of turbulence of the flow. Also, it is seen that going away from the cylinder surface, the velocity profile is deviated upward. This occurs because of the angle of the input flow into the slot which increases with augmentation of r$^*$. This can also be observable in Fig. \ref{Fig4-15}. For instance, when e$^*$=0.04 and r$^*$=0.05, the angle of the streamlines passing through the slot, is equal to $3.5^{0}$. When the slot gets away from the cylinder, this angle increases such that for r$^*$=0.5 and 0.25 it is, respectively, equal to 33$^{\circ}$ and 54.2$^{\circ}$.

\begin{figure}
	\centering
	\includegraphics[width=\columnwidth]{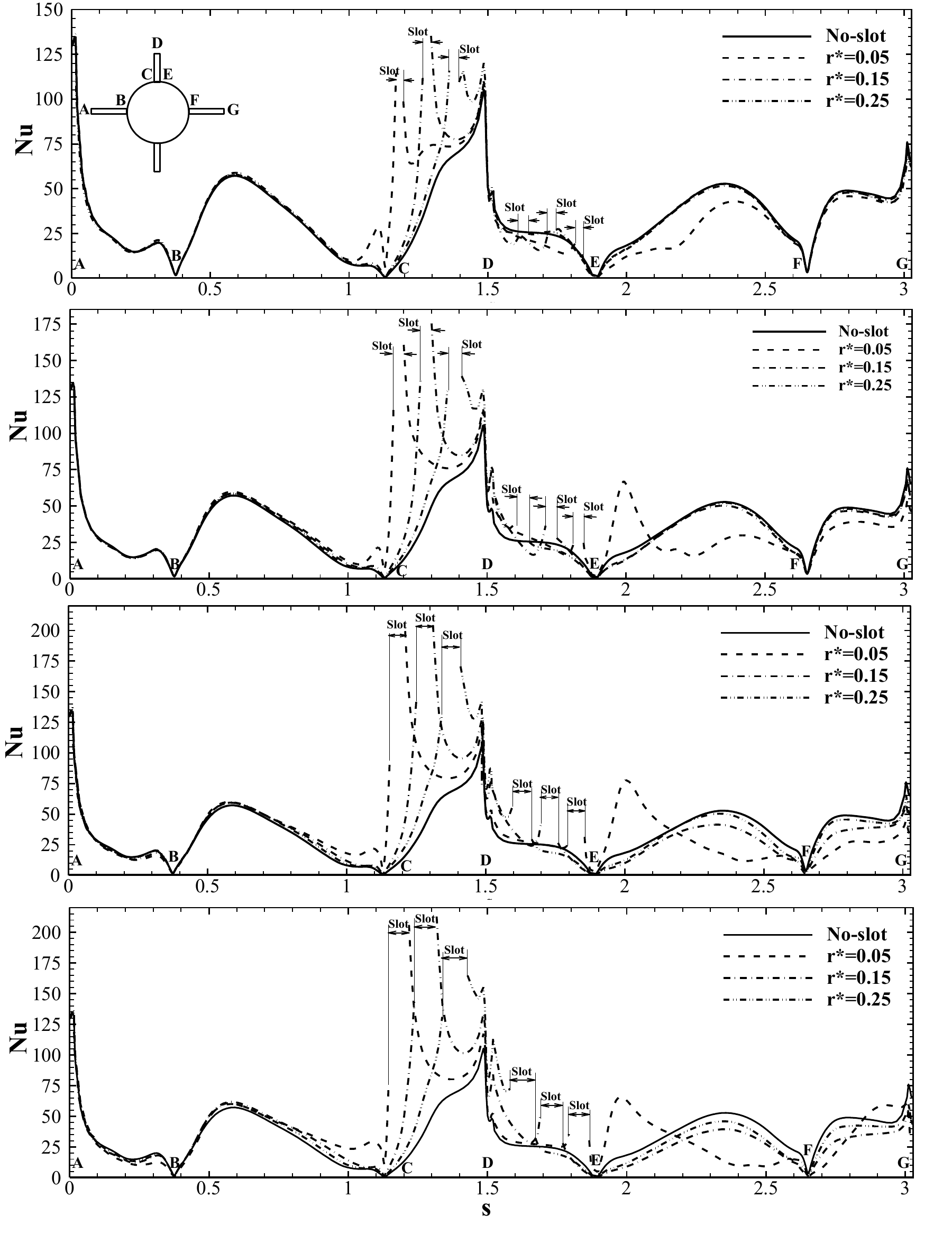}
	\caption{The local Nusselt number distribution of the cylinder with slotted fin configuration for different value of the slot distance from the surface of cylinder, r$^*$, and with the slot thickness of (a) e$^*$=0.01, (b) e$^*$=0.02, (c) e$^*$=0.04, (d) e$^*$=0.06.}
	\label{Fig4-13}
\end{figure}

Figure \ref{Fig4-13} demonstrates the effect of the slot distance from the surface of cylinder (r$^*$) on the Nusselt number distribution versus “s” for slots with different values of e$^*$. As it is distinctive, with placing the slot on the fin, no notable change of the Nu distribution from A to B corresponding to the fin$\#$1, is felt. For all e$^*$s, with increasing r$^*$, the Nu distribution over the cylinder surface from B to C tends to that of the finned cylinder without slot. The slot at r$^*$=0.05 results in Nu augmentation on the surface specially in the vicinity of Point C. The pressure difference between two sides of the slot makes suction of the low-velocity fluid from the neighborhood of the point C and its pumping to the other side of the fin$\#$2 at proximity of the point E. Thus, the dead fluid is sucked from the body surface and the fresh high-momentum flow replaces. This substitution causes the convective heat transfer enhancement and the Nu increment over the cylinder surface, and also the fin$\#$2, in closeness of the point C. In Fig. \ref{fig4-10}, as well, approaching of the streamline to the point C with insertion of the slot at r$^*$=0.15 is obvious.

The Nu distribution in the left side of the fin$\#$2, that is from point C to D; predicts as closer as the slot is to cylinder, a higher Nu near the lower region of the slot. Also, as r$^*$ increases, Nu enlarges further at the top edge of the slot but, in regions below the slot, it behaves like that of the fin without the slot. The maximum local Nu around the slot at the left side of the fin$\#$2 occurs for r$^*$=0.15. Fig. \ref{Fig4-11} also demonstrates that the value of maximum velocity for the slot placed at r$^*$=0.15 is slightly higher than velocities obtained for other two cases. The flow in the slot, is a Poiseuille-type flow which is created due to a pressure difference. The reason for this little variation of the maximum velocity is the slight discrepancy of the pressure between the two sides of the slot at different r$^*$. Section~\ref{Case1} mentions that the Nu in the middle area of the fin$\#$2 and fin$\#$4 is under the influence of the complex flow with different vortices and their interaction with each other.

\begin{figure*}
	\hspace{-1cm}
	\centering
	\includegraphics[width=12cm]{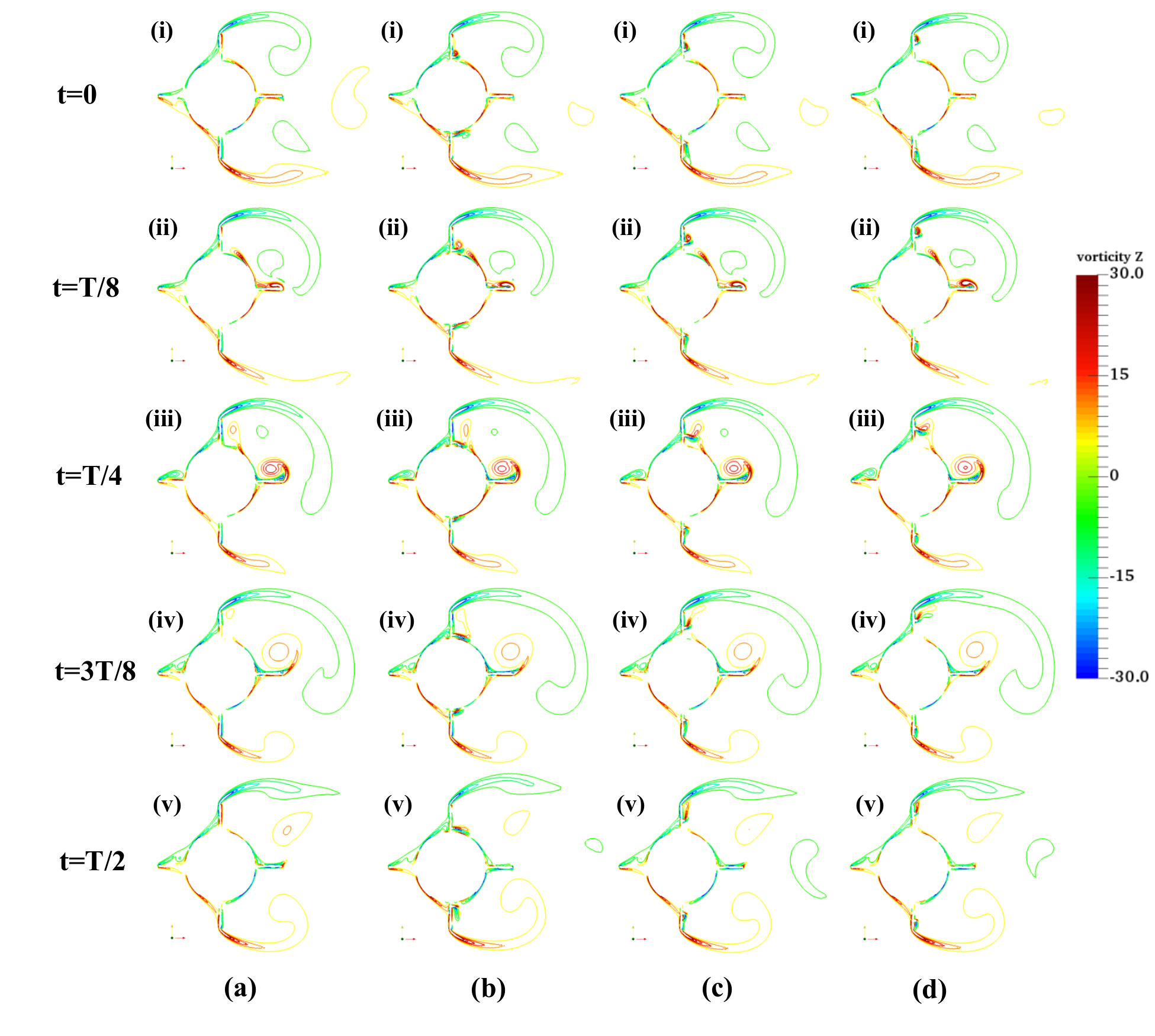}
	\caption{The vorticity contour during a half-period when e$^*$=0.01, for different values of r$^*$: (a) fined without slot, (b) r$^*$=0.05, (c) r$^*$=0.15, (d) r$^*$=0.25.}
	\label{Fig4-14}
\end{figure*}

\begin{figure*}
		\hspace{-0.5cm}
	\centering
	\includegraphics[width=11cm]{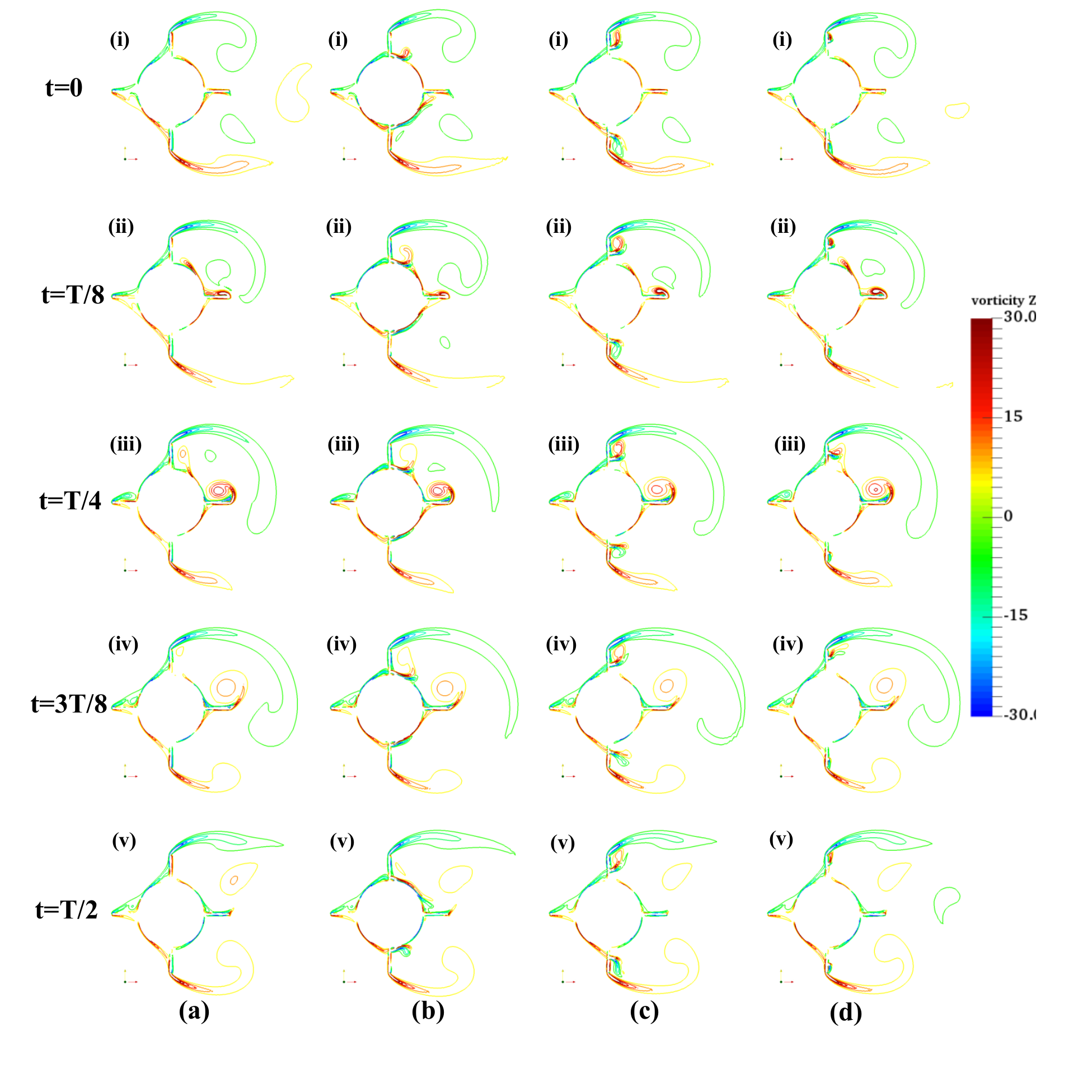}
	\caption{The same as \ref{Fig4-14} but for e$^*$=0.002.}
	\label{Fig4-15}
\end{figure*}

Fig. \ref{Fig4-14} manifests the vorticity contours at different r$^*$s for e$^*=0.01$ over a half-period, compared to the case of finned cylinder without slot. As it is seen, the output jet from the slot does not affect the flow front and the size of formation of the vortices is the same for all r$^*$. Also, it is equal to that of the finned case without slots. Fig. \ref{Fig4-14} a, which is related to the finned cylinder without slot, confirms that at point G and also on the cylinder surface, between points E and F, two vortices are simultaneously formed and grow.

The formed vortical flow, affects the vorticity field in the right side of the fin$\#$2 over the cylinder surface between the F and E points, and also on the fin$\#$4. The formed vortex at point G, grows and causes fluid mixing and convection enhancement on fin number 4 and also on the cylinder surface in the vicinity of point F. The generated vortex on the cylinder surface also grows and enhances the convective heat transfer on the cylinder surface in the neighborhood of point E and also on the fin$\#$2 surface. In general, for e$^*$=0.01, the output jet from the slot, due to its low momentum, has low penetration depth and does not have much effect on the vorticity field between the fins number 2 and 4. The output jet strikes the formed vortex and prevents it from having influence on the fin$\#$2. As this jet has no high penetration capability, does not have notable effect on increasement of the vorticity in these points. Consequently, the Nu at the right side of the fin$\#$2 and hence on the cylinder surface, specifically when r$^*$=0.05, decreases relative to the case with the non-slotted fin. Going away from the cylinder surface, the passing jet moves away from the formed vortices, and the Nusselt distribution approaches the value of Nu for a fined cylinder without slot. 

The vorticity contours over a half-period for e$^*$=0.02, at different r$^*$s, are compared with the results for a case with a non-slotted fin, in Fig. \ref{Fig4-15}. In this case, the passing jet from the slot has higher penetration power relative to the one with e$^*$=0.01. When r$^*$=0.05, it prevents the formation of vortices near the surface of the cylinder. Also, the fin$\#$2 surface and the cylinder surface in vicinity of the junction point of the fin$\#$2 to cylinder, are affected by the passing flow through the slot, and the vorticity inside these regions increases. The vorticity augmentation results in local Nu increasement, which is observable in Fig. \ref{Fig4-13}. After the passing jet flows through the slot, it detaches from the surface and reattaches the cylinder near fin$\#$4. This pattern causes a region with low Nusselt number on the cylinder surface. Hence, when the slot is in nearby the cylinder, in regions near the fin$\#$2 and fin$\#$4, respectively, Nu increase and decrease, is seen. Figure \ref{Fig4-15} indicates that when the slot is located in farther distances from the cylinder surface, the vortex formed by the flow jet, gets nearer to the fin$\#$2 surface and by increasing the vorticity, the Nu grows in the region over the slot. Moreover, when the slot keeps out from the cylinder, the effect of the passing flow through the slot, on the vortices decreases and the Nu value tends to that of the finned cylinder without slot. Additionally, Nu variations for e$^*$=0.04 and e$^*$=0.06 is also seen. As the slot widens, the penetration distance of the jet passing through the gap increases.

\subsubsection{Width of the slot}
\label{2.4.2}
\begin{figure}
	\centering
	\includegraphics[width=1.1\columnwidth]{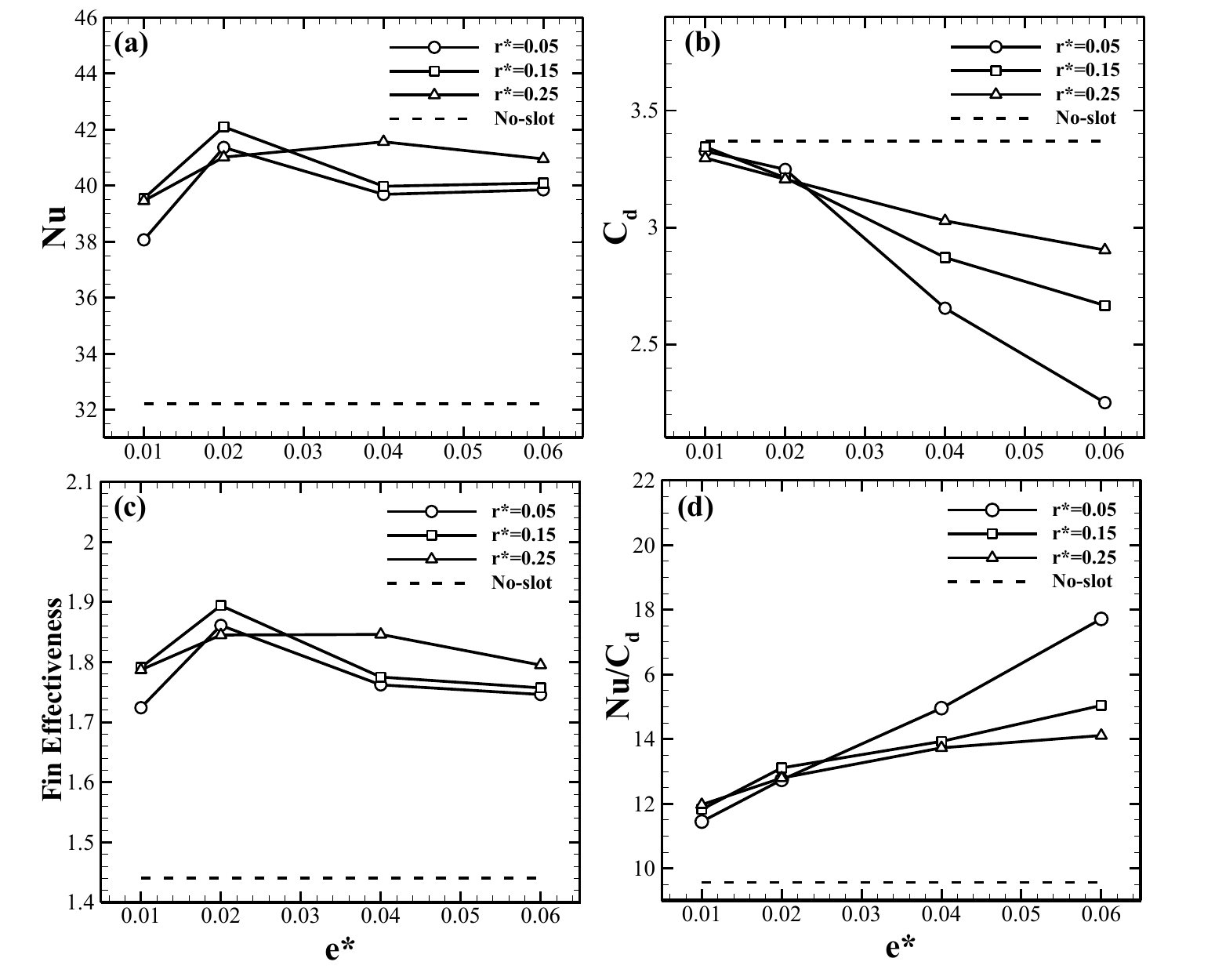}
	\caption{(a) The Nusselt number, (b) the drag coefficient, (c) the fin effectiveness, (d) Nu to drag coefficient ratio, for different slot distances from the surface, r$^*$, and also for slots with different thicknesses, e$^*$.}
	\label{Fig4-18}
\end{figure}

In Sec. \ref{2.4.1}, the effect of the slot distance from the base of the fin has been investigated. Specifically, for each width, influence of three different distances have been studied. In this section, four values of e$^*$ has been taken for each of the three contemplated r$^*$, to investigate the influence of slot width. Particularly, its effect on the drag coefficient, the Nu number, the fin effectiveness, and the Nu to drag coefficient ratio has been investigated. The impact of the slot width on various parameters has been illustrated in Fig. \ref{Fig4-18}.

Fig. \ref{Fig4-18} a presents the Nu variations versus e$^*$. As it is obvious, the behavior of the Nu number changes with increasing e$^*$, is approximately identical for both r$^*$=0.05 and r$^*$=0.15. First, the Nu augments by widening the slot. Then, increasing e$^*$ from 0.02 to 0.04, it decreases. For the case with the widest slot, the number of Nu increases slightly. For r$^*$=0.25, the longest distance between the slot and the cylinder surface, the Nu variations with increasing e$^*$ is at first increasing and then decreasing.

Fig. \ref{Fig4-18} b illustrates the drag coefficient variation versus e$^*$ for different values of r$^*$. According to this figure, for all r$^*$ values, widening the slot, causes a reduction in drag coefficient. Increasing e$^*$, the body front area reduces and also the flow rate of the high-pressure passing fluid, enlarges. This trend results in increasing the length of the vortex formation area in the downstream. With vortices getting away from the walls, the pressure on the surface increases and the form drag decreases.

The fin effectiveness versus e$^*$ which is somehow identical to the Nu variations’ dependency, is illustrated in Fig. \ref{Fig4-18} d. The fin effectiveness depends on both the Nu and the heat exchange surface between the body and the fluid; with increasing e$^*$ when t$^*$<e$^*$, or say when the width of the slot is greater than the fin thickness, the heat exchange surface reduces and with the change of e$^*$ from 0.04 to 0.06, despite the increase of the Nusselt number, the fin effectiveness has decreased.

Fig. \ref{Fig4-18} d establishes the ratio of the Nu number to the drag coefficient that increases with
e$^*$ increment for all r$^*$. As the percentage of changes of the Nu number with e$^*$ in comparison to
that of the drag coefficient, is lesser, the ratio Nu/C$_d$ is more under the influence of the drag coefficient. Due to the reduction of C$_d$, the ratio increases with the enlargement of the slot’s width. As a conclusion, Fig. \ref{Fig4-18} states that as close as the slot is to the cylinder surface, the e$^*$ dependency of the drag coefficient is more pronounced, and C$_d$ decreases with a steeper slope. Similarly, Fig. \ref{Fig4-18} emphasizes, the lesser the r$^*$ is, the ratio Nu/C$_d$ increases more with e$^*$ increasement.

\begin{figure*}
	\centering
	\includegraphics[width=1.8\columnwidth]{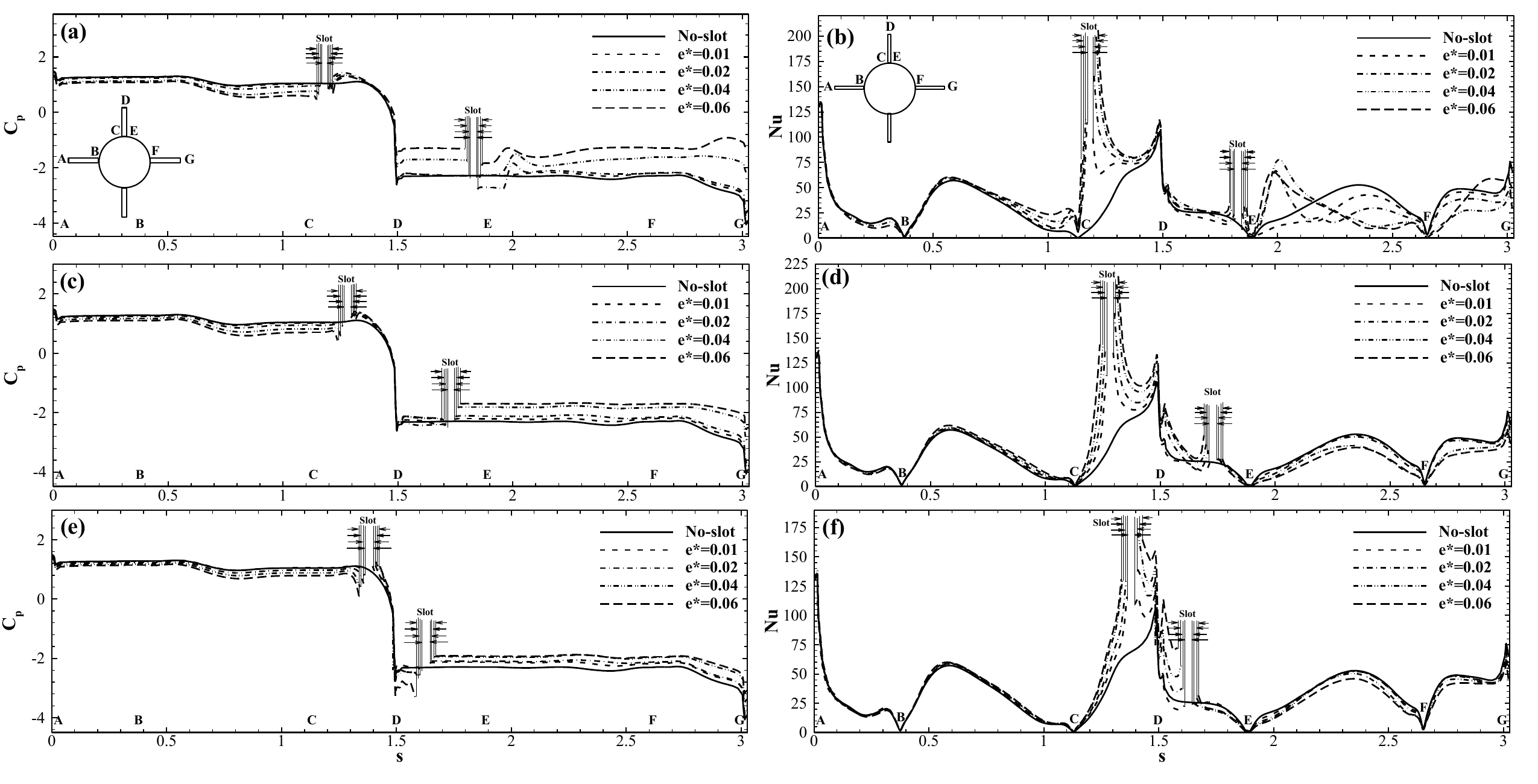}
	\caption{(a) The pressure coefficient, and (b) the local Nusselt number distributions over the cylinder with the slotted fin for different values of the slot thickness and the different slot distances from the root, (a) r$^*$=0.05, (b) r$^*$=0.15, (c) e$^*$=0.25.}
	\label{Fig4-19-20}
\end{figure*}

The pressure distribution determines the form drag coefficient and Figs. \ref{Fig4-19-20} a-f reveal the effect of the slot width on the Nusselt number distribution and the drag coefficient for different values of r$^*$. It is seen in Figs. \ref{Fig4-19-20} a, c, e that for every r$^*$, with increasing the slot width, the pressure difference between the two sides of the body decreases. This prompts the drag coefficient reduction. The closer the slot is to the cylinder, the greater the pressure coefficient variation with e$^*$ turns into. Hence, as the slot approaches surface of the cylinder, the drag coefficient dependency on e$^*$ increases. It also appears evident in Figs. \ref{Fig4-19-20} b, d, f that the effect of e$^*$ on the Nu distribution greatly differs for various parts of the body. Due to the reverse variations in diverse segments, with increasing e$^*$, the Nu variations are not monotonous and determined. From point A to B, which corresponds to fin$\#$1, the slot widening has little effect on the Nusselt number variation. Only when r$^*$=0.05, with increasing e$^*$, the Nu number decreases marginally.

%

\begin{figure}
	\hspace{-1cm}
	\centering
	\includegraphics[width=1.1\columnwidth]{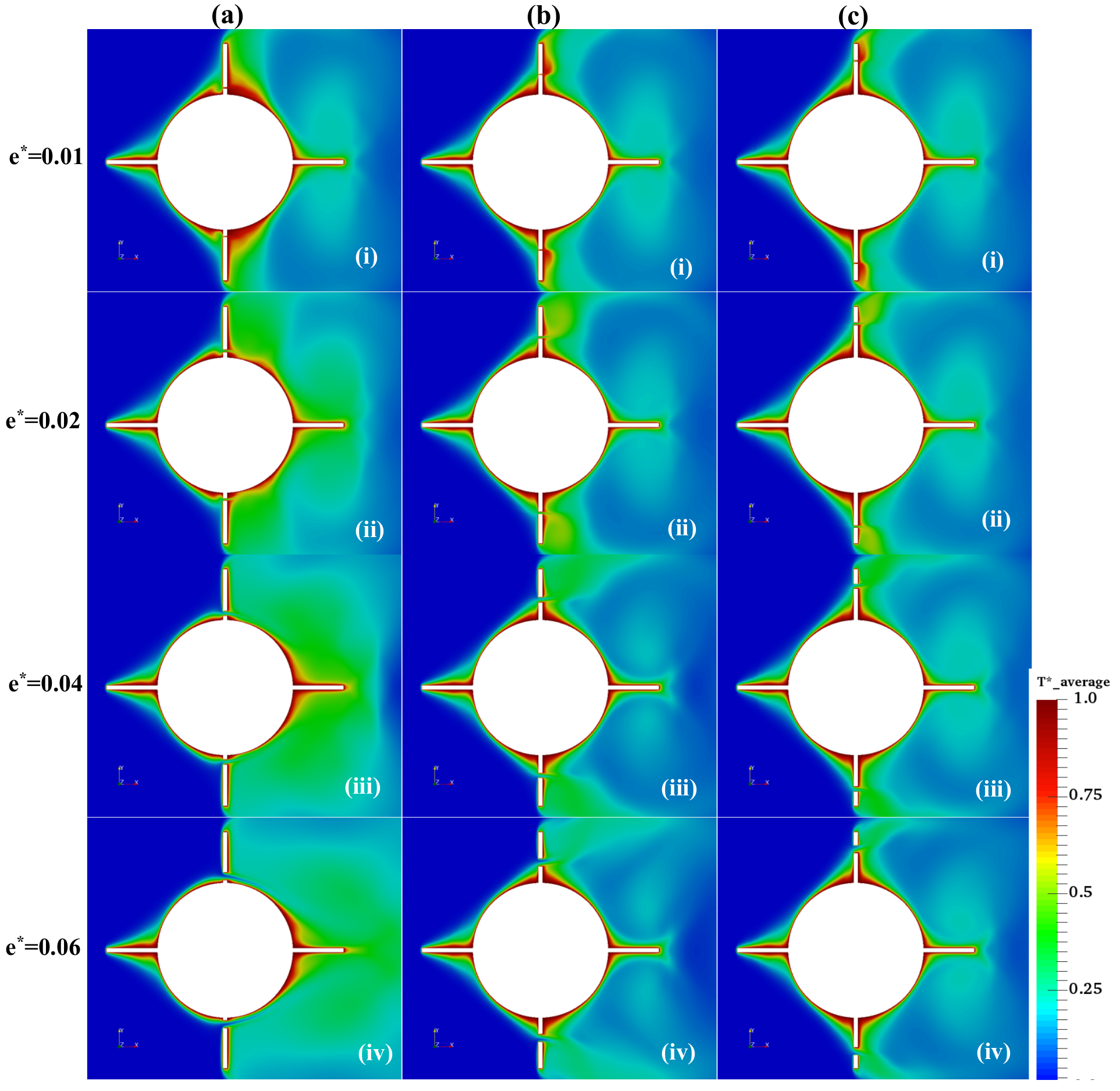}
	\caption{The time-averaged temperature field for cases with different e$^*$, and also different values of the slot distance from the surface, (a) r$^*$=0.05, (b) r$^*$=0.15, (c) r$^*$=0.25.}
	\label{Fig4-22}
\end{figure}

The time-averaged temperature field for cases with slotted fins are corroborated in Fig. \ref{Fig4-22}. When the slot is close to the cylinder surface for case a in the figure, as the slot widens, the fluid is sucked more strongly towards the slot. So, the hotter fluid has no opportunity to flow over the fin number 2, and isothermal lines are bent toward the root. This fact causes the Nusselt number reduction before point C, which is already seen in Fig. \ref{Fig4-19-20} b. For r$^*$=0.15 and r$^*$=0.25 cases that are respectively manifested in Figs. \ref{Fig4-19-20} d and f, this effect is less conspicuous and the value of Nu before point C is close to that of the finned case without slot. 

The heated flow near the surface is sucked out and a cool fluid takes its place. The wider the slot, the suction power increases, and the Nusselt number on the surface augments. The Nusselt increment over the fin when the slot is closer to the cylinder surface is larger. As the slot gets away from the surface, the value of the Nusselt number between fins$\#$2 and 4, gets closer to that of the finned cylinder without slot. Fig. \ref{Fig4-22} denotes that for all r$^*$, the wider the slot, the better the fluid at the left side of the fin$\#$2 is sucked and the cool fluid approaches it.

Also, in the right side of the fin$\#$2, it is seen that with enlarging the slot width, the leakage flow rate from the slot increases and more volume of cool high-velocity fluid is sucked into this area. Hence, the temperature of the fluid adjacent to the slot, decreases and with increasing the temperature gradient between the surface and the fluid, the Nusselt number increases. In Fig. \ref{Fig4-19-20} b, it is illustrated that at r$^*$=0.05, the Nu distribution in the region between fins$\#$2 and 4 is highly under the effect of the slot; the wider the slot, a greater Nusselt reduction occurs in vicinity of the point F. 

e$^*$ increasement results in higher penetration capability. The fluid with higher momentum flows over the cylinder surface in closeness of point E, and accordingly the Nu number increments in this region. This flow detaches from the body and reattaches to the surface of the fin$\#$4. Here, a low-vorticity region appears that weakens the convective heat transfer. Thereby, near point F, the Nu reduction is seen. The passing flow through the slot prevents vortex formation near the cylinder surface and in the widest case, affects the vortex near point G and prevents its impressment on the body. In this case, as the fluid has high momentum and attaches to the fin$\#$4 in vicinity of its end point, it increases the Nu at the end point G. Figures \ref{Fig4-19-20} d and f confirm that the Nu distribution in the region between fins$\#$2 and 4, when r$^*$=0.15 or r$^*$=0.25 is to some extent similar to that of the finned case without slot. Just a slight reduction is seen.

\subsubsection{\label{2.4.3}Number of slots}
In Secs. \ref{2.4.1} and \ref{2.4.2}, respectively, the effects of r$^*$ and e$^*$ on the drag coefficient, the Nusselt number, the Nu to C$_d$ ratio and the fin effectiveness have been investigated. In this section, in order to investigate the influence of the number of slots on the mentioned parameters, at each distance from the cylinder surface, r$^*$, the width of the slot leading to the highest fin effectiveness, is taken, in pairs for the two-slot case (N=2) and in threes for three-slot case (N=3) are placed on the fin. At r$^*$=0.05 and r$^*$=0.15, the slot with the width of e$^*$=0.02, creates the maximum effectiveness. 

In this section, the slots with e$^*$=0.02 at r$^*$=0.05, and e$^*$=0.02 at r$^*$=0.15, are named, respectively, S1 and S2. At r$^*$=0.25, the effectiveness of the fin with widths of e$^*$=0.02 and 0.04 are, respectively, 1.845 and 1.846, and both values are almost equal to each other. So, the slot width which causes a larger Nu/C$_d$ will be chosen. This ratio for e$^*$=0.02 and 0.04 is, subsequently, 12.794 and 13.728. Thus, the slot with the width of e$^*$=0.04 is selected and is assigned as S3. For N=2, three different two-slot arrangements of these three slots are contemplated. The case a is the one with S1 and S2 being both created simultaneously on the fin. It means, two slots with the width of e$^*$=0.02, are placed on the fin, at r$^*$=0.05 and r$^*$=0.15. For the case b, S1 and S3 slots are generated over the fin. The slots with e$^*$=0.02 and e$^*$=0.04 are simultaneously constructed on the fin, respectively, at r$^*$=0.05 and r$^*$=0.25. In the third two-slot fin case, called as case c, S2 and S3 are placed, consecutively, at r$^*$=0.15 and 0.25. For the three-slot case which is appointed as case d, all the three slots of S1, S2, and S3 are created concurrently on the fin. All arrangements are schematically presented in Fig. \ref{Fig11} c(i-iv).

Table \ref{Tab:4-2} demonstrates the drag coefficient, Nu, the ratio Nu/C$_d$ and the fin effectiveness for the mentioned four cases (see Figs. \ref{Fig11} c (i-iv)). Among the two-slot cases, the smallest drag coefficient belongs to the case c that has the value 16.6$\%$ less than that of the case with the non-slotted fin. Also, the maximum Nusselt number pertains to case a, which shows 55.8$\%$ increasement in comparison to the Nu for the case with the non-slotted fin. The fin with two slots in all three cases, has a higher effectiveness relative to the cases with one slot. Besides, the fin with three slots presents the maximum effectiveness. In this specimen, the heat transfer is 2.553 times the heat transfer around the cylinder without fins; while the cylinder with the non-slotted fin has increased the heat transfer in comparison to the cylinder without fin for 1.440 times. For the case d, where the three slots are simultaneously placed on the fins number 2 and 4, the drag coefficient and the Nusselt number with respect to all two-slot cases, consecutively, decreases and increase. In case d with three slots, the drag coefficient as compared to the case with non-slotted fin has been decreased for 23$\%$, and also the Nusselt number has increased for 76$\%$. The maximum ratio of the Nu/C$_d$ corresponds to the case d, with the ratio being 2.3 times larger than the value obtained for the cylinder with the non-slotted fin.

\begin{table}[htbp]
	\caption{The different case studies and the variations of related variables.}
	\label{Tab:4-2}
	\centering
	\begin{small}
		\hspace*{-0.5cm}
		\begin{tabular}{cccccc}
			\hline
			\makecell{No. of \\ Slots}  & Cases  &
			C$_d$  & Nu  & Nu/C$_d$ & $\xi$ \\ \hline
		    0 & \makecell{Fin without \\ slot}& 3.369 & 32.222 & 9.564 & 1.440 \\ \hline
		    1 & Fin with s1 & 3.248 & 41.367 & 12.736 & 1.861 \\ \hline
			1 & Fin with s2 & 3.212 & 42.102 & 13.107 & 1.894 \\ \hline
			1 & Fin with s3 & 3.028 & 41.569 & 13.728 & 1.846 \\ \hline
			2 & Case (a) & 3.046 & 50.209 & 16.483 & 2.273 \\ \hline
			2 & Case (b) & 2.841 & 49.913 & 17.568 & 2.231 \\ \hline
		\end{tabular}
	\end{small}
\end{table}

\begin{figure}
	\centering
	\includegraphics[width=\columnwidth]{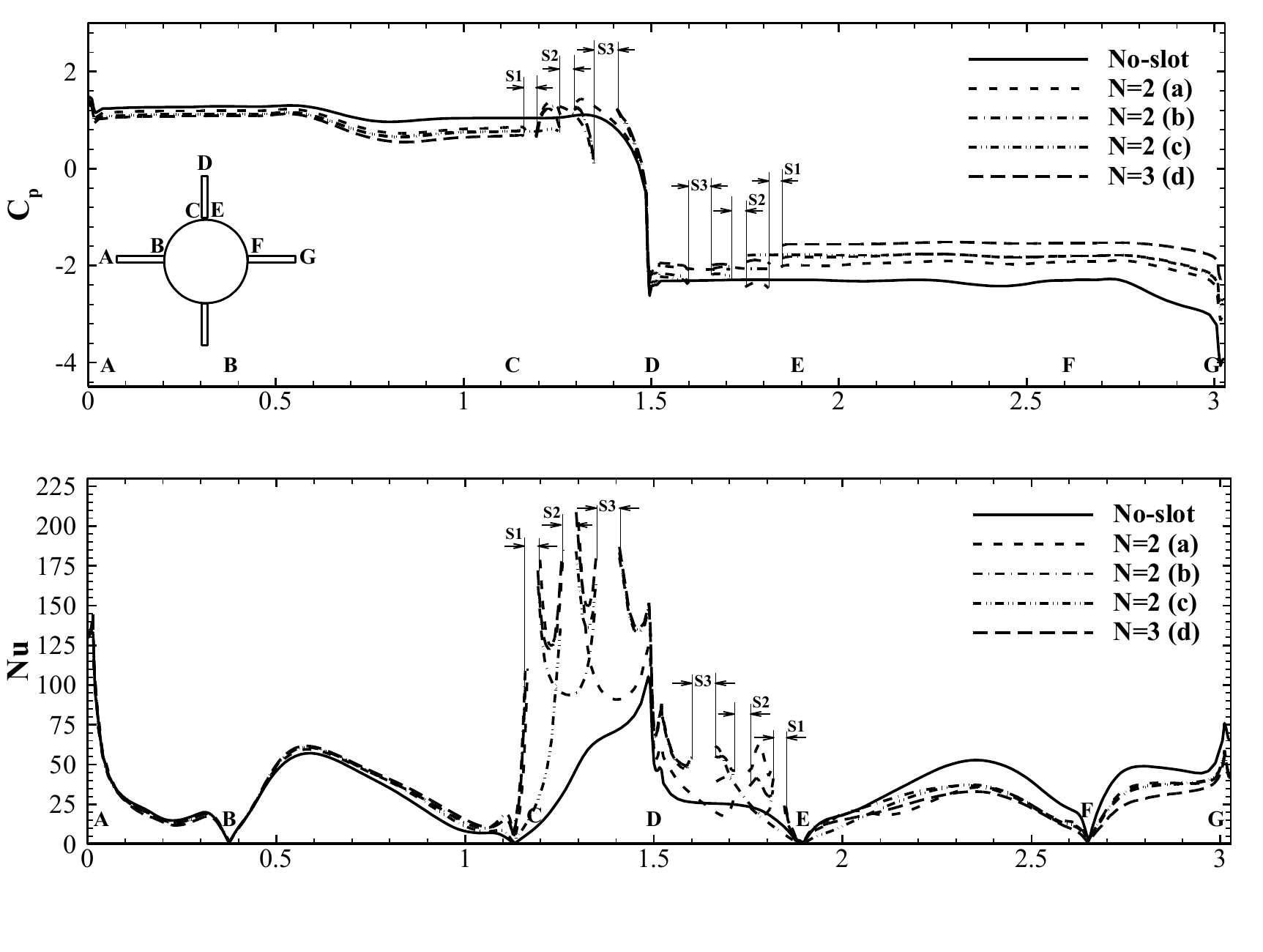}
	\caption{The local Nusselt number variation for the cylinder with slotted-fin configuration for different values of the slot distance from the surface of the cylinder, r$^*$, and with the slot thickness of: (a) e$^*$=0.01, (b) e$^*$=0.02}
	\label{Fig4-27}
\end{figure}

The pressure coefficient distribution and the local Nu variation for different two-slot cases and also the three-slotted type configuration, have been indicated in Fig. \ref{Fig4-27}. As it is conspicuous, in all cases, at the left side of the body, the pressure is reduced relative to the cylinder with non-slotted fins. On the other hand, at the right side, where the pressure has been lower, the pressure increases for all cases. Within the two-slot modes, the pressure difference diminution between the two left and right sides, for case a is lower than the two other cases and the plots for cases b and c are similar to each other. Thence, the drag coefficient of case a is larger than that of cases b and c, which only differ each other by 2.1$\%$. Fig. \ref{Fig4-27}, showing the local Nusselt number distribution for cases a to d, and the cylinder with the non-slotted fins, confirms that the main increment in Nu through the addition of a slot to the fin, takes place on the fin$\#$2 surface. It means, the Nu distribution over the fin number 2 for different two-slotted cases is a combination of two modes in which the slots are located separately on the fin. For instance, for case a, Nu at the top of S2 is similar to that of the mode in which S2 is placed solely on the fin. Also, for points located underneath S1, it is almost akin the case where S1 has been lonely placed on the fin. Between these two slots, the Nu number increases relative to the mentioned two cases.

Also, in case d that all three slots are created simultaneously on the fin, some kind of superposition occurs inside the area between the slots and, the Nu augments there. Since, the slots are placed all together, the suction of the heated fluid adjacent to the left side of the fin$\#$2, is amplified, the cool fluid is replaced, and the value of Nu rises. On the cylinder surface, between points B and C, the value of Nu for distinct cases are very close to each other. The Nu increment is not seen at vicinity of  point C for the case c. This is due to the fact that S1, the nearest slot to the cylinder surface, has not been placed on the fin for the case c. On the other side of the fin$\#$2, at the lower points of S1 and the upper points of S3, if these two slots are existent on the fin, the Nu plots of the different cases coincide. The value of Nu for this case is larger than the case without the slots. For the case d, the value of Nu in the midst of the slots S1 and S2 is more dominant than that of the case b and c, although in this region, the case a has the maximum Nu.

\subsubsection{Height of the fin}
In this section, effect of the fin height on the drag coefficient, the Nusselt number, the ratio Nu/C$_d$, and the fin effectiveness will be discussed. In all previous studies, the height of fins has been taken to be constant and equal to H=0.75. As mentioned in Sec. \ref{2.4.3}, the maximum value of the fin effectiveness and also the ratio Nu/C$_d$ occurs for the case where three slots S1, S2, S3 are simultaneously placed on the vertical fins. In this section for investigation of the effect of the height of the slotted fin, with keeping constant the value of the dimensionless height of the horizontal fins equal to H=0.75, the non-dimensional height of the fins number 2 and 3 is considered to take the values, 0.65, 0.75, 0.85. The results correspond to the cases with these three values of heights are given in Tab. \ref{Tab:4-3}.

\begin{table}[htbp]
	\caption{The calculated drag coefficient, Nusselt number, Nusselt to drag ratio, and fin effectiveness for different values of fin height, H.}
	\label{Tab:4-3}
	\centering
	\begin{small}
		\hspace*{-0.5cm}
		\begin{tabular}{ccccc}
			\hline
			\makecell{H} & \makecell{C$_d$} & Nu & Nu/C$_d$ & $\xi$ \\ \hline
			0.65 & 2.421 & 56.727 & 23.431 & 2.469 \\
			 0.75 & 2.591 & 56.738 & 21.898 & 2.553 \\
			0.85 & 2.851 & 58.743 & 20.600 & 2.656 \\ \hline
		\end{tabular}
	\end{small}
\end{table}

Obviously, increasing the height of the vertical fins, the body front’s area against the flow augments and causes a wider flow wake and increases the drag coefficient. Also as the H enlarges, the length of the region above S3, which has the highest Nusselt number increases, and consequently results in an overall Nu increment. The change of Nu with fin height augmentation from H=0.65 to H=0.75 is almost 0.02$\%$. In spite of the drag coefficient and the Nu enlargement, as the Nu variation is low relative to the drag coefficient changes, the Nu/C$_d$ ratio decreases. It is distinctive that the highest Nu/C$_d$ ratio occurs for H=0.65. Moreover, the height increment makes the fin effectiveness increase. As previously mentioned, the fin effectiveness that is proportional to the ratio of heat transfer coefficient of the finned cylinder to the non-finned cylinder, depends on both Nu and the heat exchange area. As the heat exchange area and the Nu increase with height growth, the fin effectiveness also augments. In utilization of the fin in the tube bank, the priority is given to the Nu/C$_d$ ratio and the fin with three slots and H=0.65 is chosen. Because in the tube bank, due to the closeness of tubes, when the length of the vertical fins gets longer, the fluid flow path in the lanes between the tubes becomes blocked. So, although the H increment results in fin effectiveness increasement, the shortest fin that has the highest Nu/C$_d$ ratio, is selected to be used in the bundle of tube simulations.

\subsection{Oscillating bundle of tubes with slotted fins}
\label{Bundle of tubes}
In this section, the flow and heat transfer over bundle of tubes will be studied. Specifically, using extended surfaces as a passive method and oscillation as an active method are used to investigate the heat transfer increment in an staggered orientation of bank of tubes with four columns. To fulfill this aim, first, the effect of oscillating different columns of the tube bundle will be explored, and then the consequences of adding fins to the tube bank will be discussed. According to the previous studies \cite{Williamson1988,Cheng1997,Cheng1997p2,Gau1999,Mahfouz1999,Fu2000}, when the lock-in phenomenon occurs, the heat transfer notably enhances. As aforementioned, the lock-in takes place when the excitation frequency coincides with the natural frequency. It was found~\cite{Kumar2016} that according to the amplitude-frequency plot, the lock-in does not occur for a specific frequency range, but there exists an interval in this plot, where the lock-in takes place. In fact, this plot can be divided to three parts called as the lock-in region, the transition, and the non-locked mode. As lower as the oscillation amplitude is, the lock-in phenomenon happens in a smaller interval around $f^*$=1. By augmentation of the oscillation amplitude, this interval gets larger and the lock-in becomes observable even at frequencies larger than one. 

As suggested by Kumar \emph{et al.} \cite{Kumar2016}, in order to determine in which region of the amplitude-frequency plot, the flow is placed, two conditions should be scrutinized. The first requirement is that in the power density spectrum, the dominant frequency should be equal to the excitation frequency. In other words, $f/f_0$ should be around 1 and the pick in other frequencies, should no longer exist and must be disappeared. Another condition is that the other picks in this plot, if there are any, should be an integer multiple of $f_0$ and also $f/f_0$ ratio should be an integer number. If both conditions are met, the lock-in has taken placed. If only the first condition is fulfilled, the flow is in transition or the quasi lock-in mode. If none of the conditions is satisfied, no lock-in occurs and the flow is in the non-locked zone.

\subsubsection{Bundle of fixed naked tubes}
\begin{figure}
	\centering
	\includegraphics[width=0.6\columnwidth]{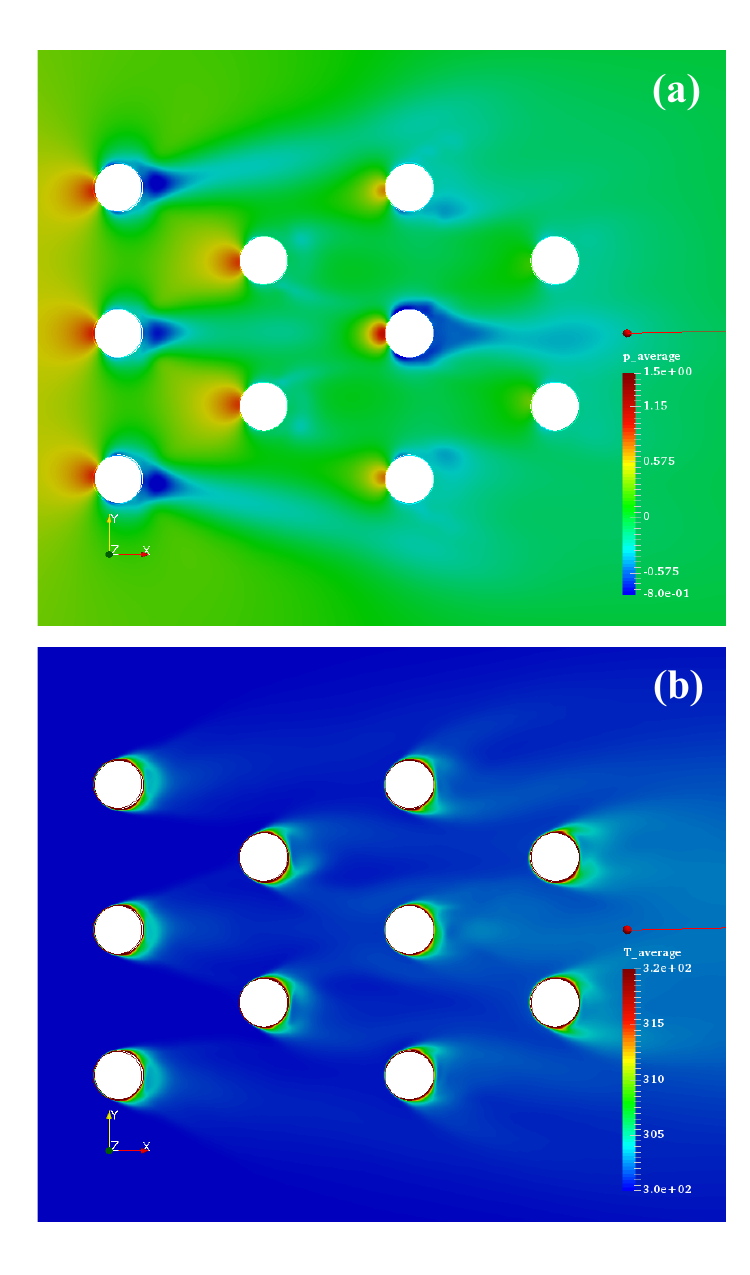}
	\caption{The time-averaged (a) pressure field and (b) temperature field around the fixed bundle of tubes without fins.}
	\label{Fig4-29}
\end{figure}

In order to find the suitable amplitude and the frequency for oscillation of different columns of the bundle of tubes, firstly, the natural frequency of the tubes in fixed state should be known. So, the flow around the fixed tube bundle at Re=5000 have been investigated. The correspondent results are presented in Tab. \ref{Tab:4-4}. As it is seen, in general, each column presents a lower drag coefficient relative to its upstream column. This drag coefficient can be attributed to the influence of the upstream tubes in the flow path, which results in formation of a low-pressure region in front of others. Also, the Nu is found to decrease for the downstream tubes owing to the fluid heating after passing the upstream tubes. Before the first column, the cool fluid hits the tubes and heats up. Figures \ref{Fig4-29} a and b present the time-averaged pressure field and temperature field, respectively, in which the pressure reduction and the temperature increment are evident. The pressure difference between two sides of the tubes around the first column is more dominant than that of the second column and so on. 

\begin{table}[htbp]
	\caption{The calculated drag coefficient, Nusselt number, Nusselt to drag ratio and Strouhal number, for each of the tubes in the naked bundle of tubes.}
	\label{Tab:4-4}
	\centering
	\begin{small}
		\hspace*{-0.5cm}
		\begin{tabular}{cccccc}
			\hline
			\makecell{Column \\ number} & \makecell{tube \\ number} & C$_d$ & Nu & Nu/C$_d$ & St \\ \hline
			1 & C$_{11}$ & 1.725 & 45.244 & 26.228 & 0.242 \\
			\cline{2-6}
			&C$_{12}$ & 1.664 & 44.984 & 27.033 & 0.212 \\ 	\cline{2-6}
			&C$_{13}$ & 1.744 & 44.912 & 25.752 & 0.242 \\ \hline
			2 & C$_{21}$ & 1.057 & 40.892 & 38.687 & 0.212 \\
			\cline{2-6}
			&C$_{22}$ & 1.077 & 41.696 & 38.715 & 0.212 \\ \hline
			3 & C$_{31}$ & 0.928 & 39.762 & 42.847 & 0.242 \\
			\cline{2-6}
			&C$_{32}$ & 1.428 & 48.137 & 33.709 & 0.212 \\ 	\cline{2-6}
			&C$_{33}$ & 0.989 & 39.962 & 40.406 & 0.242 \\ \hline
			4 & C$_{41}$ & 0.538 & 31.966 & 59.416 & 0.212 \\ \cline{2-6}
			& C$_{42}$ & 0.587 & 34.135 &58.151 & 0.212 \\ \hline
		\end{tabular}
	\end{small}
\end{table}

By moving toward the downstream columns, both the drag coefficient and the Nusselt number decrease. As the percentage of the drag coefficient changes is greater than that of the Nusselt number, the ratio of these two dimensionless numbers have been more under the influence of the drag coefficient and consequently it increases. According to the data listed in Tab. \ref{Tab:4-4}, the Strouhal number that indicates the dimensionless frequency of the vortex shedding, has two different values. In all existing tubes except the upper and the lower pipes in the first and the third columns, i.e., C$_{11}$, C$_{13}$, C$_{31}$, and C$_{33}$, St is equal to 0.212. In order for the excitation frequency of the all tubes to be close to its natural frequency, a number close to the average of these two numbers has been selected as the forced oscillation frequency. The forced frequency of the different columns of the bundle of tubes has taken to be $f_0$=0.230. In this situation, for the tubes with St=0.212, the dimensionless frequency is $f^*$=1.085, while for other tubes we have $f^*$=0.95. Both values are close to one. The ultimate goal is finding the best case from heat transfer increase and the pressure reduction point of view, by using two methods of oscillation as an active method and adding the fin as a passive method.

\subsubsection{A single oscillating slotted finned cylinder}
In the interest of finding the suitable amplitude for oscillation, a tube with the slotted fin, which in Sec. \ref{Case2} have been chosen to be utilized in a tube bank, in different amplitudes and at $f_0$=0.230 is subjected to the forced transverse oscillations. The finned cylinder oscillation have been studied for three disparate dimensionless amplitude of 0.2, 0.4, 0.6, and the results are presented in Tab. \ref{Tab:4-5}.

\begin{table}[htbp]
	\caption{The results obtained for the drag coefficient, the Nusselt number, and the Nusselt to drag ratio according to the dimensionless amplitudes for an oscillating cylinder with slotted fins; N=3 and H=0.65.}
    \label{Tab:4-5}
	\centering
	\begin{small}
		\hspace*{-0.5cm}
		\begin{tabular}{cccc}
			\hline
			\ \ \ A$^*$ \ \ \ & \ \ \ C$_d$ \ \ \ & \ \ \ Nu \ \ \ & \ \ \ Nu/C$_d$ \ \ \ \\ \hline
			\ \ \ 0.2 \ \ \ & \ 2.085 \ \ \ & \ 59.221 \ \ & \ \ 28.403 \ \ \ \\ \hline
			\ \ \ 0.4 \ \ \ & \ 2.357 \ \ \ & \ \ \ 71.150 \ \ \ & \ \ \ 30.187 \ \ \ \\ \hline
			\ \ \ 0.6 \ \ \ & \ \ 2.659 \ \ \ & \ \ \ 77.307 \ \ \ & \ \ \ 29.070 \ \ \ \\ \hline
		\end{tabular}
	\end{small}
\end{table}

\begin{figure*}
	\centering
	\includegraphics[width=1.8\columnwidth]{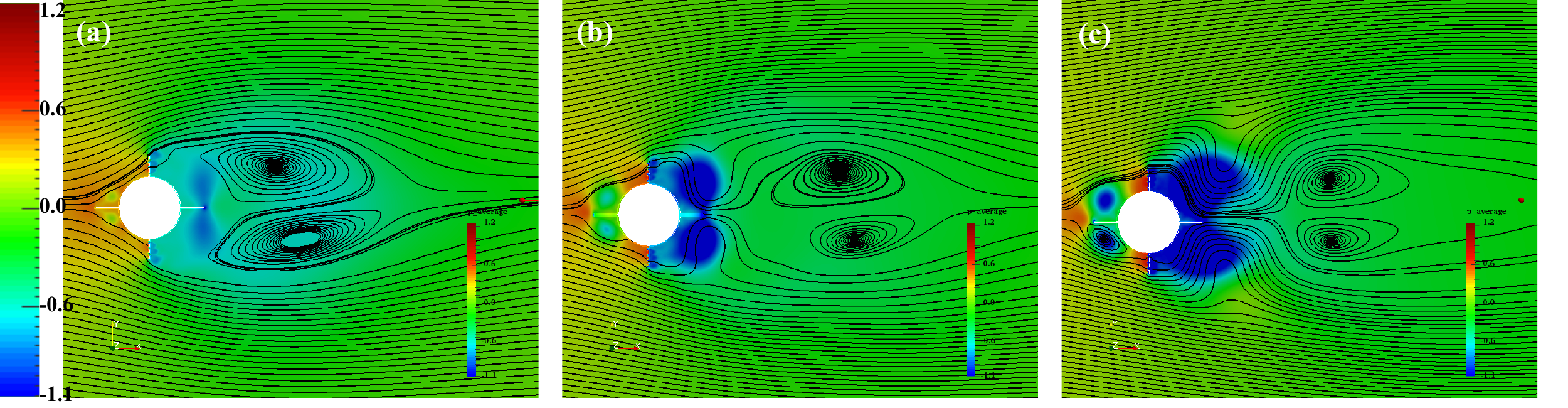}
	\caption{The pressure field and streamlines around the cylinder with slotted oscillating fin with the frequency, $f^{*}$=0.230 for different values of the oscillation amplitude: (a) A$^*$=0.2, (b) A$^*$=0.4, (c) A$^*$=0.6.}
	\label{Fig4-30}
\end{figure*}

The time-averaged pressure along with the streamlines for the cases with oscillation with three different amplitudes are demonstrated in Fig. \ref{Fig4-30}. It is seen that, as the amplitude of oscillation increases, the low-pressure zone beyond the finned cylinder expands and the pressure decreases in this region. A vortical flow appears near the tip of fin$\#$1 owing to the vertical motion of this fin inside the fluid, and a low-pressure zone is created there. The greater the amplitude of the oscillation, the lower the pressure of these vortices becomes. On the other hand, with the oscillation amplitude increment, the pressure at the left side of the fins$\#$2 and 3 increases and the pressure difference between the two sides of the finned tube augments, which causes the drag coefficient increasement. Additionally, the Nu enlarges as the oscillation amplitude rises. As the excitation frequency is the same for all three cases, the relative velocity between the fluid and the surface grows and the heat transfer rate increases with amplitude augmentation. Consequently, as the drag coefficient and the Nu, both increase with amplitude increment, and on the other hand they have a same order of magnitude variation, and due to their inverse effect on the Nu to drag ratio, the maximum of this dimensionless parameter is found for A$*$=0.4. In this circumstance, the Nu to drag coefficient ratio is increased by 28.8$\%$ relative to the fixed case.

\begin{figure}
	\centering
	\includegraphics[width=\columnwidth]{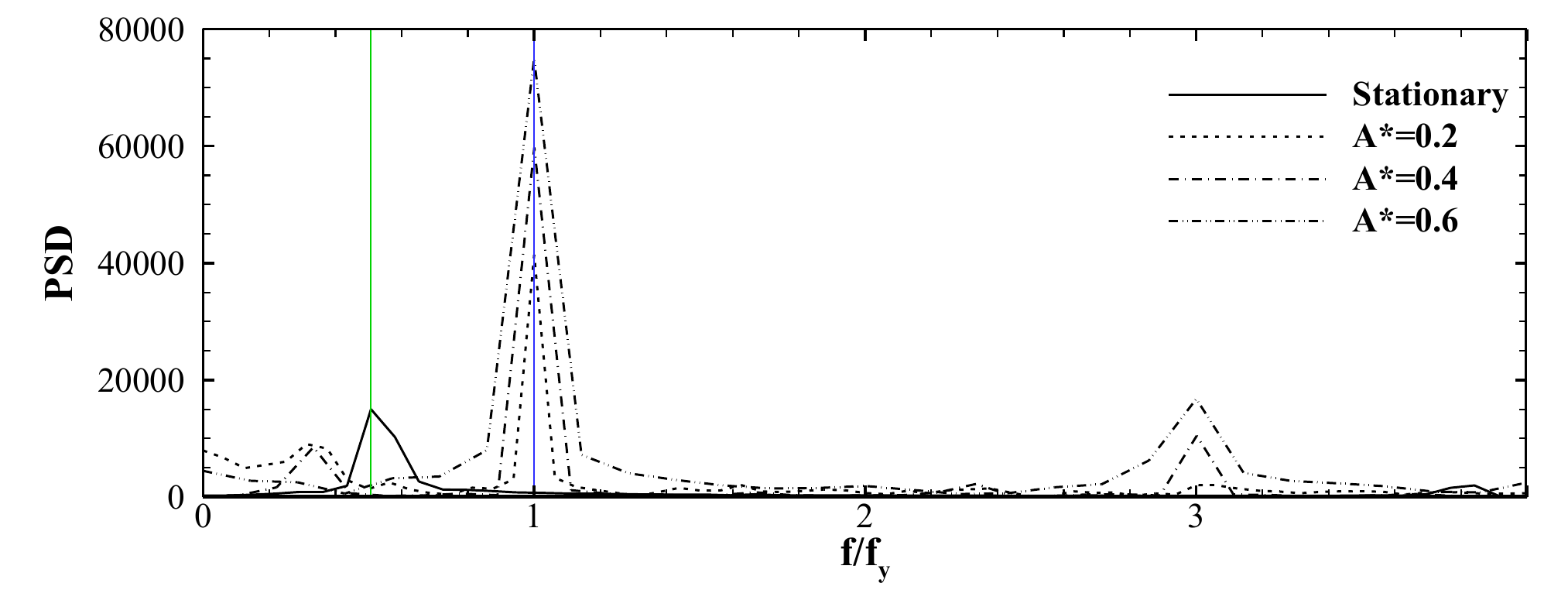}
	\caption{The power density spectrum for fixed and oscillating cylinders with the slotted fin.}
	\label{Fig4-31}
\end{figure}

Figure \ref{Fig4-31} denotes the power density spectrum for the three oscillating cases of the tube with slotted fins. For the finned tube, the dominant frequency is equal to the excitation frequency, and the natural frequency disappears. Another peak is seen at $f/f_0$=3, which is an integer multiple of the excitation frequency. It should be mentioned that for three considered oscillating amplitude, despite the fact that the excitation frequency is almost 1.8 times the natural frequency, the lock-on phenomenon takes place. As previously mentioned, Gau \emph{et al.} \cite{Gau1999} has also obtained that the lock-in, in flow around the cylinder occurs not only in frequencies equal to the excitation frequencies, but also happens when the frequency equates the integer multiple of the natural frequency of the cylinder. In the present study, it is attained that the same occurs for a finned cylinder. 

\subsubsection{Oscillating slotted-finned bundle of tubes}
\begin{figure}
	\centering
	\includegraphics[width=\columnwidth]{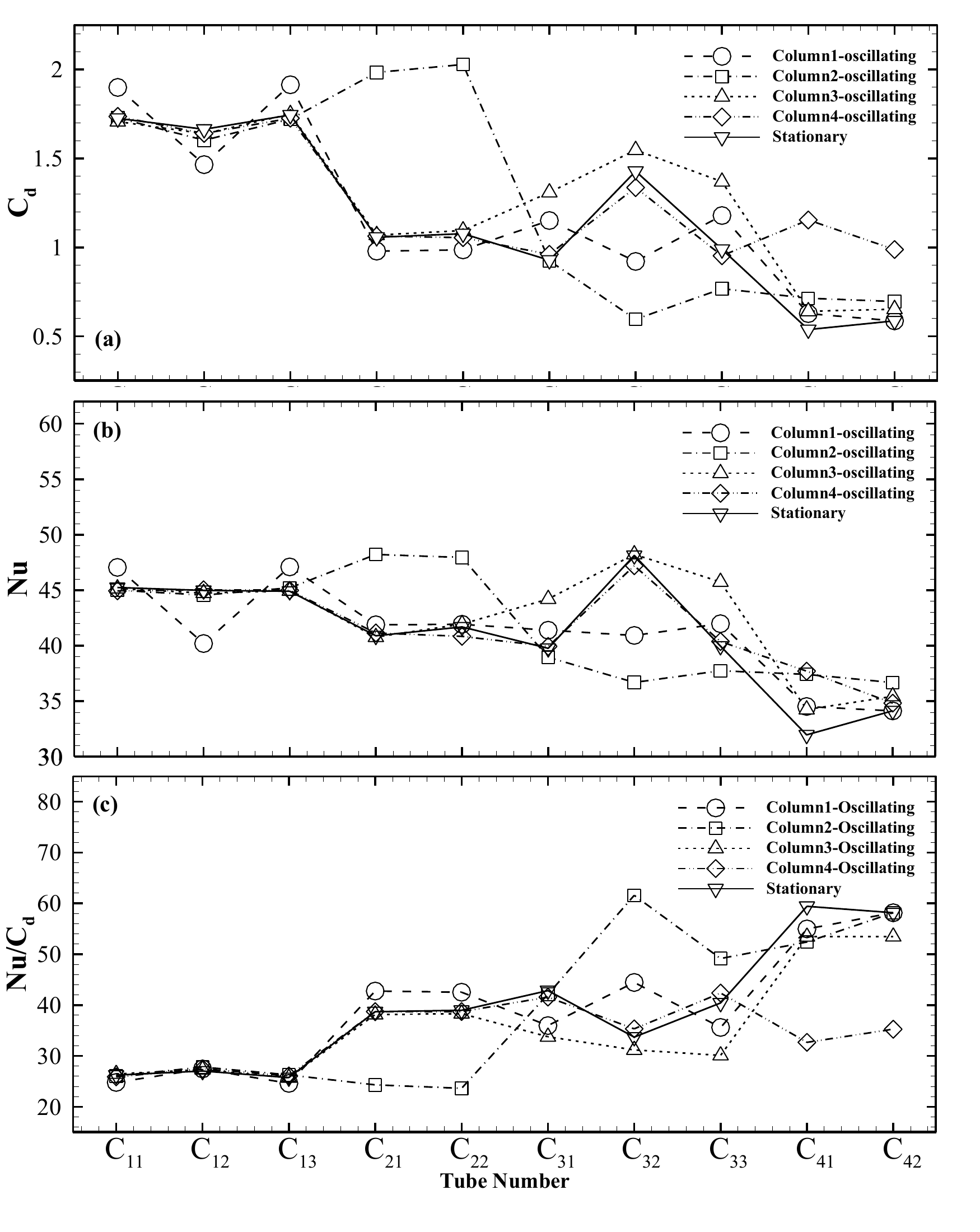}
	\caption{The effect of the position of tube in a fixed bundle of tubes with different oscillating columns on (a) the drag coefficient, (b) the Nusselt number, (c) the Nusselt number to drag coefficient ratio, (d) the fin effectiveness.}
	\label{Fig4-32}
\end{figure}

In order to investigate the effect of oscillation, different columns in the bundle of tubes with the dimensionless oscillation amplitude of A$^*$=0.4 and the frequency of $f^*$=0.230 have been subjected to the forced oscillation. The drag coefficients of the existent tubes in the tube bank versus the number of tubes in various oscillating situations are shown in Fig. \ref{Fig4-32}. Generally, for all four oscillating cases, the drag coefficient of the oscillating column is increased relative to that of the fixed case. When the first column oscillates, the drag coefficient for the mid tubes of this column with $f^*$=0.95 grows and it decreases for the mid tube with $f^*$=1.085. The same happens with the third column is oscillating, except that the drag coefficient increases for all three tubes. It is needless to say that the mid tube experiences less increment. When the second or fourth columns are oscillating, the drag coefficient of these columns enlarges in comparison to the fixed cases. In the case that the second column oscillates, the drag coefficient of the third column decreases. Oscillation of each column reduces the pressure on the low-pressure side behind it.

When the second or fourth columns are oscillating, the drag coefficient for these columns is increased compared to the stationary state. If the second column is oscillating, the drag coefficient of the third column is reduced. The pressure reduction of the behind region results in drag coefficient increment in the oscillating column. Additionally, due to the expansion of the low-pressure region behind the oscillating column, the pressure in front of the next column, placed in the downstream of the oscillating column, reduces. Accordingly, the drag coefficient of the column, next to the oscillating column, decreases. The oscillation of a column, also affects the pressure distribution of the previous column, by slightly increasing the pressure of the front side as well as the pressure at the behind of the previous one. As the changes are minor, they have no significant effect on the drag coefficient of the upstream column.

Figure \ref{Fig4-32} b compares the Nusselt number variation for different columns of four variant oscillating cases with that of the fixed cases. As it is viewed for the first and the third columns which have three tubes, in similarity to the drag coefficient, the Nusselt number for the lateral tubes with $f^*=0.95$ increases due to the oscillation. For the mid tube where $f^*=1.085$, the Nu is not augmented. In the first column, Nu of the mid tube has decreased and in the third column it is almost equal to that of the fixed one. In cases with oscillating second or forth columns, Nu augments due to the oscillation. Also, the second column oscillation results in a Nu reduction in the third column. In a general sense, the trend of Nusselt changes is similar to the drag coefficient variations. It is obtained that for the tubes with excitation frequencies less than the Strouhal frequency, the Nu increases while for the other tubes with excitation frequency larger than the Strouhal frequency, the Nusselt number either decreases or does not change relative to that of the fixed state.

Figure \ref{Fig4-32} c demonstrates the Nu to drag coefficient ratio versus the number of tubes for different oscillating cases along with the fixed state. The drag coefficient and the Nusselt number of each oscillating column augments, but, as the drag coefficient variation percentage is higher than that of the Nusselt number, the Nu/C$_d$ ratio for the oscillating column decreases. The Nu variations occurs with a smaller slope and consequently, the Nu/C$_d$ ratio follows the upward trend. Finally, what is more important is the effect of oscillation on the flow and heat transfer of the whole bundle of tubes. Thereby, parameters should be defined to present the overall performance of the tube bank for fixed and oscillating situations. The overall Nusselt and the overall drag coefficient of the bundle of tubes are defined, respectively, as the average of Nu and drag coefficient of all tubes. The ratio of overall heat transfer of the oscillating bundle of tubes to that of the fixed one, is equal to the average Nusselt ratio of all the tubes in the oscillating situation to that of the fixed state. Further, the ratio of the drag force applied on the oscillating tube bank to the force employed to the fixed case as well as the average Nu versus average drag coefficient ratio for both oscillating and fixed cases are investigated in Tab. \ref{Tab:4-6}.

\begin{table*}[htbp]
	\caption{The averaged values of parameters corresponding to the heat transfer and the resistive force for an oscillating bundle of tubes with three-slotted fins and H=0.65.}
	\label{Tab:4-6}
	\centering
	\begin{small}
		\hspace*{-0.5cm}
		\begin{tabular}{cccccc}
			\hline
			\makecell{Case studies} & Nu$_{avg}$ & C$_{d,avg}$ & $\frac{\textnormal{Nu}_{avg}}{\textnormal{C}_{d,avg}}$ & $\xi=\frac{\textnormal{Q}_{Finned}}{\textnormal{Q}_{No-Fin}}$ & $\frac{\textnormal{F}_{d,Finned}}{\textnormal{F}_{d,No-Fin}}$ \\ \hline
			\makecell{Fixed bundle of tubes} & 41.169 & 1.174 & 35.067 & - & - \\
			\hline
			\makecell{The oscillating first column} & \ \ 41.114 \ \ & \ \ 1.171 \ \ & \ \ 35.110 \ \ & 0.999 & 0.997 \\
			\hline
			\makecell{The oscillating second column} & \ \ 41.846 \ \ & \ \ 1.276 \ \ & \ \ 32.792 \ \ & 1.016 & 1.087 \\ \hline
			\makecell{The oscillating third column}  & \ \ 42.469 \ \ & \ \ 1.294 \ \ & \ \ 32.819 \ \ & 1.031 & 1.102 \\ \hline
			\makecell{The oscillating fourth column}  & \ \ 41.701 \ \ & \ \ 1.261 \ \ & \ \ 33.069 \ \ & 1.013 & 1.074 \\ \hline
		\end{tabular}
	\end{small}
\end{table*}

All oscillating cases except the first column oscillation, increase the average Nu of the tube bank in comparison to the Nu of the fixed bundle of tubes. The maximum heat transfer belongs to the third column oscillation that presents 3.1$\%$ augmentation relative to the fixed case. Also, oscillation of all columns excluding the first one, causes the applied drag force on the tube bank grow. Oscillation of the first column of the tube bank, reduces the average drag coefficient. The highest average drag coefficient occurs for the case where the third column oscillates. It shows 10.2$\%$ increment relative to the fixed one. The only oscillating case where both the Nusselt number and the drag coefficient are reduced, is the one with first column oscillating. In this situation, the maximum Nu/C$_d$ ratio is found. In other oscillating cases, the ratio Nu/C$_d$ has been decreased in comparison to that of the corresponding fixed cases. If only the heat transfer increment is under consideration, the greatest heat transfer rate takes place for the case with the third column oscillating. Normally, in order to optimize the heat exchangers, the heat transfer rate and the frictional losses are considered simultaneously. The maximum value of the Nu/C$_d$ ratio exists for the case with first column oscillating.

In the previous cases, the effect of oscillation on Nu and drag coefficient of the tube bank has been investigated. It was shown that the lock-on phenomenon occurs, for the column which is oscillating. In this section, for the sake of examining the effect of adding the chosen fin in Sec. \ref{Case1}, this fin is appended to all the tubes of the tube bank. The four preceding oscillating cases are also considered for the finned bundle of tubes. The results appeared in Tab. \ref{Tab:4-7} confirm that the maximum Nusselt of the finned oscillating bundle of tubes emerges for the case with the first column oscillating. In this case, adding fins and forcing oscillation simultaneously, increases the average Nusselt number of the bundle of tubes up to 32.5$\%$ relative to its value for the fixed tube bank.

\begin{table*}[htbp]
	\caption{The average values of the parameters corresponding to heat transport and fluid flow for an oscillating bundle of tubes with three-slotted fins and H=0.65.}
	\label{Tab:4-7}
	\centering
	\begin{small}
		\hspace*{-0.5cm}
		\begin{tabular}{cccccc}
			\hline
			\makecell{Case studies} & Nu$_{avg}$ & C$_{d,avg}$ & $\frac{\textnormal{Nu}_{avg}}{\textnormal{C}_{d,avg}}$ & $\xi=\frac{\textnormal{Q}_{Finned}}{\textnormal{Q}_{No-Fin}}$ & $\frac{\textnormal{F}_{d,Finned}}{\textnormal{F}_{d,No-Fin}}$ \\ \hline
			\makecell{Fixed bundle of tubes without fin} & 41.169 & 1.174 & 35.067 & - & - \\
			\hline
			\makecell{Bundle of pipes with fin \\ and first oscillating column} & \ \ 54.564 \ \ & \ \ 1.552 \ \ & \ \ 35.157 \ \ & 2.527 & 1.325 \\
			\hline
			\makecell{Bundle of tubes with fin \\ and second oscillating column} & \ \ 48.933 \ \ & \ \ 1.346 \ \ & \ \ 36.354 \ \ & 2.226 & 1.054 \\ \hline
			\makecell{Bundle of tubes with fin \\ and third oscillating column}  & \ \ 50.435 \ \ & \ \ 1.397 \ \ & \ \ 36.102 \ \ & 2.261 & 1.079 \\ \hline
			\makecell{Bundle of tubes with fin \\ and forth oscillating column}  & \ \ 50.673 \ \ & \ \ 1.096 \ \ & \ \ 46.234 \ \ & 2.313 & 0.869 \\ \hline
		\end{tabular}
	\end{small}
\end{table*}

For the finned oscillating tube bank in similarity to the un-finned cases, the highest drag coefficient occurs when the first column is oscillating. Besides, the minimum drag coefficient pertains to the case in which the fourth column oscillates. In this situation, the drag coefficient not only is not increased but also is decreased by 13$\%$ relative to the fixed case without the fins. It means that when the fourth column oscillates, although the front area of the tubes against the flow is enlarged, but the average drag coefficient for the overall tube bank decreases. According to the results demonstrated in Tab. \ref{Tab:4-7}, the largest heat transfer rate for the finned oscillating tube bank belongs to the case with the first column oscillating where adding the fins to the oscillating bundle of tubes increases the heat transfer rate up to 2.5 times that of the case without fins. According to the results demonstrated in Tab. \ref{Tab:4-7}, the largest heat transfer rate for the finned oscillating tube bank belongs to the case with the first column oscillating in which by adding the fin to the oscillating bundle of tubes, the heat transfer rate, compared to the case without fins increases for 2.5 times.

\section{Conclusions}
\label{con}
Turbulent forced convective heat transfer around a cylinder using the k-$\omega$ SST turbulent model is two-dimensionally simulated. The effect of adding oscillation and fins on heat transfer as an active and passive method, respectively has been investigated. Further, creating slots on the fin as a method to reduce the drag coefficient has been studied. In order to optimize the thermal systems, the heat transfer rate and the frictional losses should be contemplated concurrently. In this study, four radial fins, in pairs perpendicular to the flow and in flow direction as an extended surface are used to reinforce the heat transfer. The vertical fins make the surface of body front against the flow wider and consequently increase the low-pressure region behind the body. This fact augments the drag coefficient in comparison to that of the fixed cylinder. 

In order to reduce the drag coefficient, slots are placed on the vertical fins. The cylinder case studies with different slot distance from the cylinder surface, the width of the slot, the number of slots placed on the fins, and the fin height and oscillating frequencies are studied. The results present the drag coefficient reduction for all case studies. While the slot is thin, no notable change in pressure distribution on the surface is created and the drag coefficient does not depend much on the fin distance from the cylinder surface. 

For wider slots, the closer the slot is to the surface of the cylinder, the lower the drag coefficient becomes. The slot has a different effect on the local Nusselt number in different parts of the surface, hence, variation of the overall Nusselt number does not have a constant trend by increasing the fin distance from the cylinder. The maximum Nusselt number for a one-slotted case occurs for the one with e$^*$=0.02 and r$^*$=0.15, which shows 30.7$\%$ increase relative to that of the finned cylinder without slots. 

To find the suitable number of slots on fins, in each three considered cases, the width of slot which brings the highest fin effectiveness is chosen and in pairs or all three slots are placed on the fins. It is obtained that among the different two- and three-slotted cases, the largest Nu/C$_d$ ratio belongs to the case study with three slots. In this case, the drag coefficient and the Nusselt number, respectively, has been decreased by 23$\%$ and increased for 76$\%$ relative to the sample including the non-slotted fin. Also, the fin effectiveness has reduced from 2.553 to 1.440. The results confirm that the longer the vertical fins are, due to the greater local Nusselt number in vicinity of the end of the fin and enlargement of the heat transfer surface, the larger the overall Nu becomes. On the other hand, length increment of the fins leads to the drag coefficient growth. So, the maximum Nu to drag coefficient ratio belongs to the shortest fin.

Next, to enhance the heat transfer rate in the staggered bundle of tube configuration, first, different columns are forced to oscillate. It is acquired that the maximum Nu and the heat transfer coefficient appears when the third column is coerced to oscillate where the heat transfer increases almost by 3$\%$ relative to that of the fixed bundle of tube. Also, in all oscillating cases except the one with the first column oscillating, the drag coefficient increases. Due to the drag coefficient increment, the Nu to drag coefficient ratio decreases with oscillation. Exclusively, with the first column oscillating, the Nu/C$_d$ ratio augments for 12$\%$ in comparison to the fixed tube bank. 

Secondly, to ameliorate the heat transfer in the tube bank, the fin is added. For all tube banks with fins, the overall Nusselt number increases. The highest Nu appertains to the case with the first column oscillating. In this situation, the overall Nusselt number augments for 32.5$\%$, relative to the fixed bundle of tubes without the fin. For all the samples apart from the case with the fourth column oscillating, the fin accruing causes the drag coefficient increase. When the fourth column of the bundle of tubes oscillates, the drag coefficient decreases in comparison to both cases of oscillating and fixed bundle of tubes without fins. Consequently, the maximum Nu/C$_d$ ratio occurs for the case with oscillating fourth column, which is 31.8$\%$ higher than the one for the fixed tube bank without the fin.

In general, the oscillation alone results in the overall drag coefficient increment of the bundle of tubes and 3$\%$ increasement of the heat transfer. This is while when both oscillation and slotted fins as a combined method are applied, in spite of the drag coefficient increment, both the Nu/C$_d$ ratio and the heat transfer rate increase. At last, to optimize the functionality of oscillating heat exchangers by increasing the rate of heat transfer, and also the Nusselt number to drag coefficient ratio, adding the slotted fin is suggested.



\begin{thebibliography}{00}
	
\bibitem{Ghazanfarian2017}
J. Ghazanfarian, B. Taghilou, Active heat transfer augmentation of bundle of tubes by partial oscillatory excitation, Journal of Thermophysics and Heat Transfer, 32(3), 1-15, 2017.

\bibitem{GRIFFIN1971}
O. EM. Griffin, The unsteady wake of an oscillating cylinder at low Reynolds number, Journal of Applied Mechanics, 38, 523-532, 1971.

\bibitem{Williamson1988}
C. H. K. Williamson, A. Roshko, Vortex formation in the wake of an oscillating cylinder, Journal of fluids and structures, 2, 355-381, 1988.

\bibitem{Cheng1997}
C-H. Cheng, H-N. Chen, W. Aung, Experimental Study of the Effect of Transverse Oscillation on Convection Heat Transfer From a Circular Cylinder, Journal of Heat Transfer, 119(3): 474-482, 1997.

\bibitem{Cheng1997p2}
C-H. Cheng, J-L. HONG, Numerical prediction of lock-on effect on convective heat transfer from a transversely oscillating circular cylinder, International Journal of Heat Mass Transfer, 9310(96), 00255-4, 1997.

\bibitem{Gau1999}
C. Gau, J. M. Wu, C. Y. Liang, Heat Transfer Enhancement and Vortex Flow Structure Over a Heated Cylinder Oscillating in the Crossflow Direction, Journal of Heat Transfer, 121(4), 789-795, 1999.

\bibitem{Mahfouz1999}
F.M. Mahfouz, H.M. Badr, Forced convection from a rotationally oscillating cylinder placed in a uniform stream, International Journal of Heat and Mass Transfer, 43, 3093-3104, 1999.

\bibitem{Fu2000}
W-S Fu , B-H. Tong, Numerical investigation of heat transfer from a heated oscillating cylinder in a cross flow, International Journal of Heat and Mass Transfer, 45(14), 3033-3043, 2000.

\bibitem{Pottebaum2006}
T.S. Pottebaum, M. Gharib, Using oscillations to enhance heat transfer for a circular cylinder, International Journal of Heat and Mass Transfer, 49(17-18), 3190-3210, 2006.

\bibitem{Kumar2016}
S. Kumar, Navrose, S. Mittal, “Lock-in in forced vibration of a circular cylinder”, Physics of Fluids, 28(11), 113605, 2016.

\bibitem{Mousavi2018}
S.B. Mousavi, M.M. Heyhat, Numerical study of heat transfer enhancement from a heated circular cylinder by using nanofluid and transverse oscillation, Journal of Thermal Analysis and Calorimetry, 135, 935–945, 2018.

\bibitem{Murthy1983}
J. Y. Murthy, S. V. Patankar, Numerical Study of HEat Transfer From a Rotating Cylinder with External Longitudinal Fins, Numerical Heat Transfer, 1983, 6(4), 463-473.

\bibitem{Abu-Hijleh2003}
B.A/K. Abu-Hijleh, Numerical simulation of forced convection heat transfer from cylinder with high conductivity radial fins in cross-flow, International Journal of Thermal Sciences, 42(8), 741-748, 2003.

\bibitem{Abu-Hijleh2003p2}
B.A/K. Abu-Hijleh, Enhanced Forced Convection Heat Transfer From a Cylinder Using Permeable Fins, Journal of Heat Transfer, 125(5), 804-811, 2003.

\bibitem{Rahnama2004}
M. Rahnama, M.Farhadi, Effect of radial fins on two-dimensional turbulent natural convection in a horizontal annulus, International Journal of Thermal Sciences, 43(3), 255-264, 2004.

\bibitem{Haldar2004}
S. C. Haldar, Laminar Free Convection Around a Horizontal Cylinder with External Longitudinal Fins, Heat Transfer Engineering, 25(6), 45–53, 2004.

\bibitem{Haldar2007}
S.C. Haldar, G.S. Kochhar, K. Manohar, R.K. Sahoob, Numerical study of laminar free convection about a horizontal cylinder with longitudinal fins of finite thickness, International Journal of Thermal Sciences, 46(7), 692-698, 2007.

\bibitem{Khashehchi2014}
M. Khashehchi, I. Ashtiani Abdi, K. Hooman, T. Roesgen, A comparison between the wake behind finned and foamed circular cylinders in cross-flow, Experimental Thermal and Fluid Science, 52, 328-338, 2014.

\bibitem{McClure2016}
J. McClure, S. Yarusevych, Vortex shedding and structural loading characteristics of finned cylinders, Journal of Fluids and Structures, 65, 138-154, 2016.

\bibitem{Bouzari2016}
S. Bouzari, J. Ghazanfarian, Unsteady Forced Convection over Cylinder with Radial Fins in Cross Flow, Applied Thermal Engineering, 112, 214-225, 2016.

\bibitem{Islam2020}
M.R. Islam , and A. Mohany, Vortex shedding characteristics in the wake of circular finned cylinders, Physics of Fluids, 32(4), 2020.

\bibitem{Tsutsui2002}
T. Tsutsui, T. Igarashi, Drag reduction of a circular cylinder in an air-stream, Journal of Wind Engineering, 90(4-5), 527-541, 2002.

\bibitem{Wang2006}
J.J. Wang, P.F. Zhang, S.F. Lu, K. Wu, Drag Reduction of a Circular Cylinder Using an Upstream Rod, Flow, Turbulence and Combustion, 76, 83–101, 2006.

\bibitem{Hwang2007}
J-Y Hwang, K-S Yang, Drag reduction on a circular cylinder using dual detached splitter plates, Journal of Wind Engineering and Industrial Aerodynamics, 95(7), 551-564, 2007.

\bibitem{Qiu2014}
Y. Qiu, Y. Sun, Y.Wu, Y. Tamura, Effects of splitter plates and Reynolds number on the aerodynamic loads acting on a circular cylinder, Journal of Wind Engineering and Industrial Aerodynamics, 127, 40-50, 2014.
\bibitem{Klausmann2017}

K. Klausmann, B. Ruck, Drag reduction of circular cylinders by porous coating on the leeward side, J. Fluid Mech, 813, 382-411, 2017.

\bibitem{Amiraslanpour2017}
M. Amiraslanpour, J. Ghazanfarian, S.E. Razavi, Drag suppression for 2D oscillating cylinder with various arrangement of splitters at Re=100: A high-amplitude study with OpenFOAM, Journal of Wind Engineering and Industrial Aerodynamics, 164, 128-137, 2017.

\bibitem{Ghazanfarian2015}
J. Ghazanfarian, R. Saghatchi, M Gorji-Bandpy, Turbulent Fluid-Structure Interaction of Water-Entry/Exit of a Rotating Circular Cylinder using SPH Method, International Journal of Modern Physics C, 26(8), 1550088, 2015.

\bibitem{Ghazanfarian2016}		
J. Ghazanfarian, R. Saghatchi, M. Gorji-Bandpy, SPH simulation of Turbulent Flow past a High-Frequency In-line Oscillating Cylinder near Free-Surface, International Journal of Modern Physics C, 27(12), 1650152, 2016.

\bibitem{Ghazanfarian2019}
J. Ghazanfarian, Introducing the Sliding-Wall Concept for Heat Transfer Augmentation: The Case of Flow over Square Cylinder at Incidence, Heat Transfer Engineering, 41, 751-764, 2020.

\bibitem{Rajani2016}
B.N. Rajani, A. Kandasamy, S. Majumdar, LES of Flow past Circular Cylinder at Re=390, Journal of Applied Fluid Mechanics, 9(3), 1421-1435, 2016.

\bibitem{Bouhairie2007}
S. Bouhairie, V.H. Chu, Two dimensional simulation of unsteady heat transfer from a circular cylinder in crossflow, Journal of Fluid Mechanics, 570, 177–215, 2007.

\bibitem{Norberg1987}
C. Norberg, Effects of Reynolds number and a low-intensity free stream turbulence on the flow around a circular cylinder, Chalmers University, Goteborg, Sweden, Technological Publications, 87(2), 1-55, 1987.

\bibitem{Mittal2001}
S. Mittal, Computation of three-dimensional flows past circular cylinder of low aspect ratio, Physics of Fluids, 13, 177, 2001.

\bibitem{Tamura1990}
T. Tamura, On the reliability of two-dimensional simulation for unsteady flows around a cylinder-type structure, Journal of Wind Engineering and Industrial Aerodynamics, 35, 275-298, 1990.

\bibitem{Schmidt1943}
E. Schmidt, K. Wenner, Heat transfer over the circumference of a heated cylinder in transverse flow, Tech. Rep. 1050. Nat. Advisory Cmte Aero (NACA), 12(2), 1943.

\bibitem{Tutar2000}
M. Tutar, A.E. Hold, Large Eddy Simulation of a Smooth Circular Cylinder Oscillating Normal to a Uniform Flow, Journal of Fluids Engineering, 122(4), 694-702, 2000.

\bibitem{Zdanski2016}
P.S. Zdanski, Jr.M. Vaz, and G.T. Gargioni, Convection Heat Transfer Enhancement on Recirculating Flows in a Backward Facing Step: The Effects of a Small Square Turbulence Promoter, Heat Transfer Engineering, 37(2),162–171, 2016.


\end{thebibliography}
\end{document}